\documentclass{iucrjournals}

\usepackage{graphicx} 
\usepackage{amsmath}
\usepackage{amsfonts}

\usepackage{acronym}
\usepackage{hyperref}
\usepackage{subfig}
\usepackage{float}
\usepackage{amssymb}
\usepackage{multirow}

\usepackage{caption}
\title{Stochastic Optimisation for Scalable Ptychographic Image Reconstruction}

\author[a,b]{A. W. Muriithi}%
\author[a]{Jared Tanner}%
\author[a,d]{Jaroslav M. Fowkes }
\author[c]{Benedikt J. Daurer\IUCrAufn{Corresponding Author}}%

\affil[a]{Mathematical Institute, University of Oxford, Oxford, UK }
\affil[b]{Ada Lovelace Center, Science and Technology Facilities Council (STFC), Rutherford Appleton Laboratory, Didcot, UK }
\affil[c]{Diamond Light Source Ltd., Harwell Science and Innovation Campus, Didcot, Oxfordshire, UK }
\affil[d]{Computational Mathematics Theme, Science and Technology Facilities Council (STFC), Rutherford Appleton Laboratory, Didcot, UK }

\renewcommand{\harvardurl}[1]{\url{#1}}

\makeatletter
\renewcommand{\harvarditem}[4][\null]{\item[]%
\if@filesw{ \def\protect##1{\string ##1\space}%
\ifthenelse{\equal{#1}{\null}}
  {\def\next{{#4}{#2}{#2}{#3}}}
  {\def\next{{#4}{#2}{#1}{#3}}}
\immediate\write\@auxout{\string\harvardcite\codeof\next}%
}\fi%
\protect\hspace*{-\labelwidth}\protect\hspace*{-\labelsep}\hypertarget{cite.#4}{}\ignorespaces%
}
\renewcommand{\HAR@dolist}[2]{\def\@citea{\null}\@for\@citeb:=#1\do%
{\@citea\def\@citea{\HAR@hisep\penalty\@m\ }\HAR@checkdef{\@citeb}%
{\hyperlink{cite.\@citeb}{#2{\@citeb}\HAR@hysep\ \HAR@year{\@citeb}}\HAR@setd{\@citeb}}}%
}
\makeatother

\begin{document}

\maketitle

\begin{abstract}

X-ray ptychography is a lensless coherent diffraction imaging technique capable of high spatial resolution, in which an image is reconstructed by solving a large-scale inverse problem from highly redundant far-field diffraction measurements acquired under overlapping illumination. Whilst the measurement redundancy created by overlapping illumination is essential for reconstruction fidelity, it leads to computational and memory demands that scale poorly with object size and target resolution when using conventional reconstruction algorithms.

This work proposes a ptychographic reconstruction framework that reduces peak memory usage and time to convergence by leveraging stochastic optimisation techniques developed for large-scale machine learning. Specifically, the framework employs stochastic gradient descent with a decaying step size that progressively reduces the effect of gradient noise, referred to here as variance reduction (VR), with gradients that agree with automatic differentiation, to optimise the ptychographic forward model using randomly sampled subsets of diffraction measurements. For an experimental dataset of 200,000 diffraction patterns of $128\times128$ pixels, VR is demonstrated to require $65\%$ of the peak GPU memory and $22\%$ of the reconstruction time of the established maximum-likelihood (ML) algorithm.

The VR framework incorporates the Poisson maximum-likelihood formulation and partial coherence within a unified and extensible optimisation setting. Numerical experiments demonstrate that the framework achieves reconstruction quality visually comparable to established methods, with lower peak GPU memory and a shorter time to convergence than the maximum-likelihood algorithm in the configurations tested.

\end{abstract}

\section{Introduction}
\label{sec:Introduction}

X-ray ptychography is a lensless imaging technique which uses coherent radiation to resolve fine-scale structure with high spatial resolution, without the use of focusing optics~\cite{Rodenburg_2025}. An image is then recovered from the resulting diffraction data by solving a large-scale inverse problem based on redundant far-field measurements acquired under overlapping illumination~\cite{Rodenburg_2004}.

The reconstruction problem in ptychography is inherently high-dimensional and involves the joint recovery of the object transmission function, the illuminating probe, and, in practical settings, the probe positions. A wide range of iterative algorithms have been developed to address this problem, differing in how they enforce physical constraints, model experimental noise, and trade off computational cost against reconstruction fidelity. While these methods have enabled routine high-resolution imaging, they exhibit significant limitations in terms of scalability, robustness to experimental imperfections, and memory efficiency when applied to modern large-scale datasets~\cite{Nashed_2017, Dong_2018, Yu_2021}.

In particular, many established reconstruction approaches either rely on sequential local updates, which limit parallel efficiency~\cite{Enders_2016}, or require global operations over all diffraction patterns~\cite{Nashed_2014}, leading to high per-iteration computational and memory costs. The computational and memory challenges associated with ptychographic reconstruction are further amplified in realistic experimental regimes, including partially coherent illumination~\cite{Yu_2016}, probe instability~\cite{Dwivedi_2017}, and three-dimensional extensions such as ptycho-tomography~\cite{Grsoy_2017}. Due to these computational, memory, and modelling challenges, there remains a need for reconstruction algorithms that reduce computational burden while maintaining robustness and modelling flexibility.

\subsection{Object and Probe Recovery}

Early ptychographic reconstruction methods alternate between enforcing the overlap and modulus constraints. The Ptychographical Iterative Engine (PIE), introduced by Rodenburg \cite{Rodenburg_2004}, applies these constraints sequentially to recover an object estimate under the assumption of a known probe. PIE updates local regions of the object corresponding to individual probe positions, making each update computationally inexpensive but resulting in serial object updates.

Thibault \textit{et al.} introduced the Difference Map (DM) algorithm to jointly recover both probe and object \cite{Thibault_2008,Thibault_2009}. DM alternates projections onto constraint sets and uses their difference to form updates, enforcing global consistency across all probe positions. Compared to the sequential local-update strategy used by PIE, DM evaluates all scan positions during each iteration. This increases the per-iteration computational cost and memory usage but improving robustness to probe uncertainty~\cite{Thibault_2009} and reducing the time to iterate through the dataset.
The extended PIE (ePIE) algorithm removed the requirement for a known probe by alternating object and probe updates using local information \cite{Maiden_2009}. ePIE uses one diffraction frame to calculate probe and object updates at each iteration, whereas DM uses all diffraction frames to calculate an update at each iteration. Whilst ePIE has a reduced per-iteration cost relative to DM, it remains sensitive to noise and experimental imperfections and often requires careful tuning of update parameters~\cite{Moshtaghpour_2025_LoRePIE}.

In contrast, statistical modelling approaches cast ptychographic reconstruction as a likelihood maximisation problem over the experimental measurements. Thibault and Guizar-Sicairos formulated a maximum-likelihood (ML) objective based on Poisson statistics of photon-counting detectors \cite{Thibault_2012}. The maximum-likelihood formulation provides a principled treatment of noise but also incurs significant computational cost due to the need to evaluate gradients over all diffraction patterns. The maximum-likelihood framework, DM and ePIE algorithms were subsequently extended to a partially coherent illumination model where the probe is represented as a sum of incoherent modes \cite{Thibault_2013}.
Other ptychographic reconstruction schemes include Relaxed Averaged Alternating Reflections (RAAR) \cite{Luke_2004}, semi-implicit Douglas–Rachford (sDR) splitting \cite{Pham_2019}, least-squares ML (LSQ-ML) \cite{Odstril_2018}, and ADMM-based formulations \cite{Yan_2019}.

\paragraph{Computational Scaling Challenges}
Ptychography requires the acquisition and processing of large numbers of diffraction patterns, and typically reconstruction costs scale unfavourably with detector size, object size, and number of probe positions~\cite{Dong_2018}. The unfavourable scaling of reconstruction cost is exacerbated in three-dimensional extensions such as ptycho-tomography \cite{Dierolf_2010}, where memory requirements and computational cost grow rapidly with sample volume and number of projections.
Algorithms that process all diffraction patterns simultaneously require the entire dataset to be loaded into memory for each update, with some also requiring the storage of intermediate wavefields, gradients, and auxiliary variables. This imposes substantial memory demands and tightly couples the processing of neighbouring views, complicating parallelisation. Whilst GPU acceleration and multi-node parallelism reduce wall-clock time \cite{Enders_2016,Favre_2020,Wakonig_2020,Yu_2021}, these hardware-based acceleration strategies do not fundamentally improve algorithmic scalability and often require access to large GPU clusters with high memory capacity \cite{Dong_2018,Yu_2021}. The memory requirements, parallelisation constraints, and dependence on large GPU resources motivate the development of reconstruction algorithms that explicitly reduce per-iteration cost and peak memory usage.

This work proposes a ptychographic reconstruction framework that reduces per-iteration cost and peak memory usage, whilst maintaining reconstruction accuracy, by combining small fixed-size mini-batches with a decaying step-size schedule~\cite{Robbins_1951,Bottou_2016} that progressively reduces the effect of gradient noise; we refer to this as variance reduction (VR).

The contributions of this work are as follows:
\begin{enumerate}
\item A mini-batch reconstruction algorithm for ptychography that minimises the Poisson negative log-likelihood with a multi-mode probe model, samples diffraction patterns in proportion to their intensity, and reduces the effect of gradient noise with a decaying step size (Section~\ref{sec:Problem_Definition}).
\item Time and memory models for this algorithm and for the full-batch ML algorithm of \texttt{PtyPy}, with constants measured by independent benchmarks (Section~\ref{subsec:Computational_Complexity_Analysis}).
\item A comparison with the DM, ePIE and ML algorithms of \texttt{PtyPy} on three experimental datasets from Diamond Light Source, of up to $2\times10^{5}$ diffraction patterns (Section~\ref{sec:Experimental_Demonstration}).
\item Synthetic scaling experiments of up to $1.02\times10^{6}$ diffraction patterns that test the time and memory models (Section~\ref{sec:Simulation}).
\end{enumerate}
Section~\ref{sec:Experimental_Setup} describes the hardware, the reporting metrics and the convergence criterion, and Section~\ref{sec:Conclusion} summarises the findings and the questions they leave open.

\section{Problem Definition}
\label{sec:Problem_Definition}

In a ptychography experiment, a coherent X-ray beam, referred to as the probe $\boldsymbol{P}$, illuminates an object $\boldsymbol{O}$, commonly modelled as a complex-valued transmission function. The interaction between the probe and object produces an exit wave-field, $\boldsymbol{\psi}$, that encodes information about the object structure.

At a given probe position, the exit wave is modelled discretely as the element-wise product of the spatially discretised probe and object,
\begin{equation}
\label{Wavefield equation}
\boldsymbol{\psi}_{i,j} = \boldsymbol{P}_{i,j}(\boldsymbol{r}) \cdot \boldsymbol{O}_{i,j}(\boldsymbol{r}),
\end{equation}
where $(i,j)$ index spatial coordinates in the object plane.
To acquire spatially localised information, the probe is scanned across the object and multiple diffraction patterns are recorded. At the $k^{th}$ scan position, the probe is displaced by a transverse shift $\boldsymbol{r}_k$, resulting in the exit wave
\begin{equation}
\label{Overlap Constraint}
\boldsymbol{\psi}_{i,j}(\boldsymbol{r}_k) = \boldsymbol{P}_{i,j}(\boldsymbol{r}-\boldsymbol{r}_k) \cdot \boldsymbol{O}_{i,j}(\boldsymbol{r}).
\end{equation}
The scan positions are chosen to introduce overlap between adjacent illuminated regions of the object. The overlap between adjacent illuminated regions introduces redundancy in the measurements, which is central to successful ptychographic reconstruction.
Each exit wave propagates from the object plane to the detector, where an intensity-only diffraction pattern $\boldsymbol{I}$ is measured. The detected intensity is modelled as
\begin{equation}
\label{Intensity Equation}
\boldsymbol{I}_{i,j,k} = \left| \mathcal{P}\left[\boldsymbol{\psi}_{i,j}(\boldsymbol{r}_k)\right] \right|^2,
\end{equation}
where $\mathcal{P}$ denotes the wave propagation operator from the object plane to the detector plane. In the far-field regime, the propagation operator $\mathcal{P}$ is commonly approximated by a discrete Fourier transform $\mathcal{F}$, up to a known complex-valued scaling factor~\cite{Kappeler_2017}. Since only intensities are recorded at the detector and phase information is lost, the measurement process gives rise to the phase retrieval problem.
The ptychographic reconstruction task consists of recovering the probe $\boldsymbol{P}$ and object $\boldsymbol{O}$ from a collection of intensity measurements acquired at different probe positions. This can be expressed as the nonlinear inverse problem
\begin{equation}
\label{Ptychography inverse problem}
\boldsymbol{I}_{i,j,k} = \left| \mathcal{F}\left[\boldsymbol{\psi}_{i,j}(\boldsymbol{r}_k)\right] \right|^2 + \epsilon,
\end{equation}
where $\epsilon$ represents measurement noise, which is well modelled as Poisson-distributed for photon-counting detectors~\cite{Thibault_2012}.
Each exit wave corresponds to a limited field of view determined by the probe support, such that the spatial dimensions of $\boldsymbol{I}$ and $\boldsymbol{P}$ are smaller than those of the reconstructed object $\boldsymbol{O}$. Consequently, the inverse problem in Equation~\ref{Ptychography inverse problem} is ill-posed and admits non-unique solutions~\cite{Fann_2020}. Ambiguities arise from, for example, an arbitrary complex scaling between $\boldsymbol{P}$ and $\boldsymbol{O}$, as well as global phase offsets. Moreover, modelling errors and experimental imperfections generally preclude an exact solution consistent with all measurements.
To render the problem tractable, additional constraints derived from physical modelling and experimental design are imposed~\cite{Thibault_2009}. A common constraint enforces consistency with the measured diffraction intensities via the modulus constraint,
\begin{equation}
\label{Modulus Constraint}
\left| \mathcal{F}\left[\boldsymbol{P}_{i,j}(\boldsymbol{r}-\boldsymbol{r}_k) \cdot \boldsymbol{O}_{i,j}(\boldsymbol{r}) \right] \right|
= \sqrt{\boldsymbol{I}_{i,j,k}}.
\end{equation}
A second key constraint arises from the intentional overlap between adjacent probe positions as formulated in Eq.~\ref{Overlap Constraint}, which introduces spatial redundancy into the diffraction data and is essential for resolving phase ambiguities and enabling stable reconstruction.
From an optimisation perspective, ptychography can be formulated as estimating $\boldsymbol{P}$ and $\boldsymbol{O}$ that best explain the measured diffraction data under the forward model. 

Given the measured intensities $\boldsymbol{I}$ and estimated probe positions $\boldsymbol{r}_k$, recovery of the probe and object is commonly achieved by minimizing a loss function that quantifies the discrepancy between measured and predicted intensities. The resulting optimisation problem can be written as
\begin{equation}
\label{Ptychography Optimisation Problem}
\boldsymbol{P}, \boldsymbol{O} \approx
\underset{\hat{\boldsymbol{P}}, \hat{\boldsymbol{O}}}{\mathrm{argmin}}
\sum_{k=1}^{K}
\mathcal{L}\left(
\hat{\boldsymbol{I}}_k(\hat{\boldsymbol{P}}, \hat{\boldsymbol{O}}),
\boldsymbol{I}_{k}
\right).
\end{equation}
Here, $\mathcal{L}(\cdot, \cdot)$ denotes a loss function quantifying mismatch with measured data, and $\hat{\boldsymbol{P}}$ and $\hat{\boldsymbol{O}}$ are estimates of the unknown parameters which are used to calculate an estimate of a given diffraction pattern, $\hat{\boldsymbol{I}}_k$. Due to noise and model mismatch, the optimal solution to Eq.~\ref{Ptychography Optimisation Problem} generally corresponds to a non-zero loss when reconstructing experimental data.

\subsection{Multi-Modal Reconstruction and Diffraction Pattern Scaling}
\label{Multi-Modal Reconstruction}

The assumption of a fully coherent probe is a simplifying approximation, as X-ray sources generally exhibit partial spatial and temporal coherence~\cite{Vartanyants_2010}. Partial coherence can be modelled by reconstructing multiple incoherent probe modes~\cite{Konijnenberg_2017}. In this work, we adopt the multi-modal formulation proposed by Thibault and Menzel~\cite{Thibault_2013}.
Specifically, the probe is represented as a set of $M$ modes. For a given probe position, the exit waves corresponding to each mode are propagated independently and the resulting intensities are summed incoherently. The resulting modulus constraint is given by
\begin{equation}
\label{Scaling and Incoherent Modes}
\sum_{m=1}^{M}
\kappa_{k} \left|
\mathcal{F}
\left[
\boldsymbol{P}_{i,j,m}(\boldsymbol{r}-\boldsymbol{r}_k)\cdot\, \boldsymbol{O}_{i,j}(\boldsymbol{r})
\right]
\right|^2
=
\boldsymbol{I}_{i,j,k},
\end{equation}
where $\kappa_{k} \in \mathbb{R}_+^1$ is a per-pattern scaling factor that accounts for fluctuations in probe amplitude across the dataset. The per-pattern scaling parameters $\kappa_k$ are optimized jointly with the probe and object.

\subsection{Loss Function}
\label{Loss Function Details}

Diffraction patterns are typically recorded using photon-counting detectors, for which measurement noise is well described by Poisson statistics~\cite{Thibault_2012}. While the Poisson distribution approaches a Gaussian distribution in the high-count regime, the Gaussian approximation becomes inaccurate for low-dose measurements.
In the Poisson noise model, for predicted intensities $\hat{\boldsymbol{I}}_k = \boldsymbol{I}(\hat{\boldsymbol{P}}, \hat{\boldsymbol{O}}, \boldsymbol{r}_k)$, the loss is given by
\begin{equation}
\label{PoissonNLL Equation}
\mathcal{L}(\hat{\boldsymbol{I}}_k, \boldsymbol{I}_{i,j,k})
=
\hat{\boldsymbol{I}}_k
-
\boldsymbol{I}_{i,j,k}\,\log(\hat{\boldsymbol{I}}_k)
+
\log(\boldsymbol{I}_{i,j,k}!),
\end{equation}
where all operations are applied element-wise.
Set-projection algorithms such as DM implicitly assume that the constraint sets intersect, an assumption often violated in experimental data, leading to stagnation or instability~\cite{Thibault_2012}. Likelihood maximisation algorithms, such as our proposed VR algorithm do not make the same assumption however. Moreover, additional physical effects can be incorporated straightforwardly by augmenting the loss function, provided the forward model remains differentiable.
An advantage of likelihood-based optimisation over set-projection algorithms is that noise and model mismatch are naturally incorporated into the objective function.
By default, PtyPy assumes a Gaussian noise model and minimizes the Gaussian negative log-likelihood,
\begin{equation}
\label{GaussianNLL Equation}
\mathcal{L}(\hat{\boldsymbol{I}}_k, \boldsymbol{I}_{i,j,k})
=
\frac{1}{2\,\boldsymbol{\sigma}_{i,j,k}^2}
\left(
\hat{\boldsymbol{I}}_k - \boldsymbol{I}_{i,j,k}
\right)^2,
\end{equation}
where the spatially varying variance is approximated as~\cite{Thibault_2012,Ptypy}
\begin{equation}
\label{GaussianNLL Sigma Estimation Equation}
\boldsymbol{\sigma}_{i,j,k}^2 \approx \boldsymbol{I}_{i,j,k} + 1.
\end{equation}
To ensure a fair comparison in Sections~\ref{sec:Experimental_Demonstration} and~\ref{sec:Simulation}, reconstructions in VR are performed twice for each dataset using a fixed random seed: once to record runtime and memory usage, and once to compute Gaussian NLL values for direct comparison with PtyPy.

\subsection{Probe initialisation}
\label{Probe Initialisation}

In many experimental settings, a reasonable estimate of the probe is available, for example from prior measurements at the same instrument. In this work, we restrict attention to datasets for which an approximate probe model is provided and use this as the initial probe estimate.
To initialise a multi-modal probe when only a single-mode probe estimate is available, we compute a singular value decomposition (SVD) of the single-mode probe estimate,
\begin{equation}
\label{Model Probe SVD}
\bar{\boldsymbol{P}}_{0,i,j} = \boldsymbol{U} \boldsymbol{\Sigma} \boldsymbol{V}^{\mathsf{T}},
\end{equation}
and retain the $M$ dominant components to initialize the probe modes,
\begin{equation}
\label{Model Probe SVD Truncation}
\bar{\boldsymbol{P}}_{0,i,j,m} = \boldsymbol{\sigma}_m \, \boldsymbol{u}_m \boldsymbol{v}_m^{\mathsf{T}},
\quad m = 1,\dots,M.
\end{equation}
Each probe mode is normalized to unit mean magnitude,
\begin{equation}
\label{Initial probe scaling}
\hat{\boldsymbol{P}}_{0,i,j,m}
=
\frac{i'j'\,\bar{\boldsymbol{P}}_{0,i,j,m}}
{\sum_{i=1}^{i'}\sum_{j=1}^{j'} |\bar{\boldsymbol{P}}_{0,i,j,m}|},
\end{equation}
where $i'$ and $j'$ denote the probe pixel dimensions. Normalising each probe mode to unit mean magnitude improves numerical stability and enables consistent optimizer hyperparameters across datasets.

\subsection{Optimisation Framework}
\label{Optimisation Framework}

Ptychographic reconstruction may be formulated as an optimisation problem by defining an objective function that measures the discrepancy between measured diffraction intensities and those predicted by a candidate object–probe pair as described in Eq.~\ref{Ptychography Optimisation Problem}. Minimizers of this objective function are taken as solutions to the ptychographic inverse problem.

\paragraph{Gradient Descent}
Gradient descent is a first-order iterative method that updates parameters in the direction of steepest descent of the objective function. The update rule is

\begin{equation}
\boldsymbol{\theta}_{l+1} = \boldsymbol{\theta}_l - \boldsymbol{\alpha}_l \frac{1}{K} \sum_{k=1}^{K}\nabla_{\boldsymbol{\theta}} \mathcal{L}(\hat{\boldsymbol{I}}(\boldsymbol{\theta}_{l,k}), \boldsymbol{I}_{k}).
\label{Gradient Descent Step}
\end{equation}

where $\boldsymbol{\theta}$ denotes the optimisation variables (the real-space pixel values of the object $\boldsymbol{O}$ and probe $\boldsymbol{P}$, and the per-pattern scaling factors $\kappa_k$), $l$ is the iteration index, and $\boldsymbol{\alpha}$ is the step size. The choice of $\boldsymbol{\alpha}$ strongly influences convergence behaviour.

Automatic differentiation frameworks, such as PyTorch \cite{Paszke_2019}, enable efficient computation of gradients by representing the forward ptychographic model as a computational graph of differentiable operations \cite{Nashed_2017}. This allows gradients of the loss function to be computed directly with respect to each object and probe pixel using only first-order information. 
Whilst gradient descent is straightforward to implement, it is computationally inefficient for large ptychographic datasets and often exhibits slow or suboptimal convergence \cite{Li_2023}, motivating the use of stochastic gradient descent.

\paragraph{Stochastic Gradient Descent}
Stochastic gradient descent (SGD) reduces the cost of evaluating the gradient update over an entire dataset by approximating the gradient using a subset of the data at each iteration,
\begin{equation}
\boldsymbol{\theta}_{l+1} = \boldsymbol{\theta}_l - \boldsymbol{\alpha} \frac{1}{K'} \sum_{k=1}^{K'}\nabla_{\boldsymbol{\theta}} \mathcal{L}(\hat{\boldsymbol{I}}(\boldsymbol{\theta}_{l,k}), \boldsymbol{I}_{k}),
\quad K' < K.
\label{Stochastic Gradient Descent Step}
\end{equation}

Here, $K'$ denotes the number of diffraction patterns used to form the gradient estimate. The subset of size $K'$ is termed a mini-batch, and the full dataset $K$ is the batch.
Mini-batch gradient estimates introduce variance relative to the full-batch gradient \cite{Johnson_2013}, but substantially reduce per-iteration cost and have been shown to converge efficiently in large-scale inverse problems \cite{Barbieri_2025}.

Selecting mini batches using uniform random sampling yields unbiased gradient estimates \cite{Chen_2019}, but does not account for variation in information content across diffraction patterns. Patterns with higher total intensity typically contain stronger signal and greater probe information \cite{Du_2019}. We therefore adopt intensity-weighted sampling, assigning selection probability
\begin{equation}
\label{Sampling Bias}
\Pi(k) = \frac{\|\boldsymbol{I}_k\|_F^2}{\sum_{n=1}^{K}\|\boldsymbol{I}_n\|_F^2},
\end{equation}
where $\|\cdot\|_F$ denotes the Frobenius norm. Sampling is performed with replacement, such that the probability of selecting a given diffraction pattern is independent across mini-batches. High-intensity frames are therefore selected more often, while every frame with non-zero intensity keeps a non-zero probability of selection, so all frames can contribute to gradient updates during a reconstruction.

For the reconstructions reported here, the gradient of the Poisson negative log-likelihood with respect to the object and probe is evaluated with a hand-derived Wirtinger expression rather than by back-propagation through the computational graph. The hand-derived gradient agrees with the automatic-differentiation gradient to floating-point rounding, but avoids storing the intermediate quantities of the forward model, which lowers the peak memory of each update.

\paragraph{Adaptive Optimiser}
In this work, we employ the NAdam optimizer \cite{Dozat_2016}, which combines adaptive per-parameter learning rates with Nesterov momentum to improve convergence speed and stability. Adaptive learning rates mitigate the imbalance in convergence rates between low- and high-spatial-frequency components observed in maximum-likelihood ptychography updates \cite{Thibault_2012}.

\subsection{Variance Reduction}

The method used here reduces the effect of gradient noise on the iterates by decreasing the step size. Throughout this paper, variance reduction (VR) refers to this mechanism.

In the experiments reported in Sections~\ref{sec:Experimental_Demonstration} and~\ref{sec:Simulation} the mini-batch size $K'$ is held fixed throughout each reconstruction, resulting in reduced memory complexity compared to variance reduction with an increasing batch size. In the general variance-reduction framework however, the mini-batch size can be adjusted dynamically during optimisation. The step size used in gradient descent, $\boldsymbol \alpha$, also plays a critical role in convergence behaviour. Excessively large values amplify gradient noise and can lead to instability, while overly small values slow convergence \cite{Smith_2020,Wang_2021}. In this work, $\boldsymbol \alpha$ is decreased over successive iterations to satisfy the Robbins–Monro conditions \cite{Robbins_1951,Bottou_2016}.

To make the variance-reduction mechanism explicit, the mini-batch gradient estimator $g_l$ may be decomposed as
\begin{equation}
g_l
=
\nabla \mathcal{L}(\boldsymbol{\theta}_l)
+
\xi_l,
\qquad
\mathbb{E}[\xi_l \mid \boldsymbol{\theta}_l]=0,
\end{equation}
where $\nabla \mathcal{L}(\boldsymbol{\theta}_l)$ is the full-batch gradient and $\xi_l$ represents the stochastic error introduced by unbiased mini-batch sampling. The SGD update can therefore be written as
\begin{equation}
\boldsymbol{\theta}_{l+1}
=
\boldsymbol{\theta}_l
-
\alpha_l \nabla \mathcal{L}(\boldsymbol{\theta}_l)
-
\alpha_l \xi_l.
\end{equation}

Under standard bounded-variance assumptions, the variance of the stochastic perturbation satisfies
\begin{equation}
\mathbb{E}\!\left[\left\|\xi_l\right\|^2\right]
=
\mathcal{O}\!\left(\frac{\alpha_l^2}{K'_l}\right),
\end{equation}
where $K'_l$ denotes the mini-batch size at iteration $l$. For our VR algorithm, $K'$ is independent of iteration number. Consequently, the stochastic component decreases as the mini-batch size increases or as the step size decreases. In the limiting case of full-batch gradient descent ($K'_l=K$), the sampling error vanishes ($\xi_l=0$), while for fixed mini-batch size the influence of the stochastic term tends to zero as $\boldsymbol{\alpha}_l \rightarrow 0$.

Consequently, the SGD trajectory approaches that of full-batch gradient descent as either the mini-batch size increases or the step size decreases. In the present work, variance reduction is obtained primarily through a decaying step-size schedule, which progressively lowers the stochastic noise floor while retaining the computational advantages of small mini-batches. The decaying step-size schedule enables exploratory noisy gradient updates in early iterations, and more stable refinement later \cite{Smith_2017,Li_2014}. Furthermore, maintaining a small, constant mini-batch size avoids adverse optimisation behaviour associated with large-batch stochastic optimisation \cite{Sclocchi_2023,Arnaboldi_2024}. 

\subsection{Computational Complexity Analysis}
\label{subsec:Computational_Complexity_Analysis}

\newcommand{\SynAMLTPassErr}{+30.6}
\newcommand{\SynAMLMemErr}{-1.5}
\newcommand{\SynAMLL}{5855}
\newcommand{\SynAVRTPassErr}{-28.3}
\newcommand{\SynAVRMemErr}{-12.6}
\newcommand{\SynAVRL}{266}
\newcommand{\SynBMLTPassErr}{-7.8}
\newcommand{\SynBMLMemErr}{-0.3}
\newcommand{\SynBMLL}{11630}
\newcommand{\SynBVRTPassErr}{+0.3}
\newcommand{\SynBVRMemErr}{-7.1}
\newcommand{\SynBVRL}{315}
\newcommand{\SynCMLTPassErr}{-2.0}
\newcommand{\SynCMLMemErr}{-0.4}
\newcommand{\SynCMLL}{11169}
\newcommand{\SynCVRTPassErr}{-2.8}
\newcommand{\SynCVRMemErr}{-5.1}
\newcommand{\SynCVRL}{338}
\newcommand{\SynDVRTPassErr}{-2.5}
\newcommand{\SynDVRMemErr}{-1.8}
\newcommand{\SynDVRL}{405}
\newcommand{\LargestVREpochUpdateShare}{9}
\newcommand{\LargestVREpochFFTShare}{91}
\newcommand{\IFourteenMLTIterErr}{-0.9}
\newcommand{\IFourteenVRTEpochErr}{-24.8}
\newcommand{\SynDMLMemPred}{142}
\newcommand{\SynDMLDataPred}{134}

\paragraph{Computational Complexity of the VR Algorithm Proposed}
Mini-batch stochastic optimisation enables gradient updates to be approximated using subsets of the diffraction data, thereby reducing both computational cost and memory requirements. We analyse the complexity of the proposed algorithm with respect to dataset size and optimiser, retaining the leading-order terms the respective multiplicative constants. Every constant is either a byte count read from the respective implementation, or a quantity measured by a benchmark script given in the code accompanying the paper. The scaling variables used throughout this analysis are summarised in Tab.~\ref{Scaling parameters table}.

\begin{table}[H]
\centering
\caption{Definition of parameters used in the asymptotic computational complexity analysis.}
\label{Scaling parameters table}
\begin{tabular}{|c|c|}
\hline
\textbf{Parameter} & \textbf{Description} \\
\hline
$K'$ & Number of diffraction patterns per mini-batch \\
\hline
$K$ & Total number of diffraction patterns in the dataset \\
\hline
$N_{\mathrm{object}}$ & Number of object pixels \\
\hline
$N_{\mathrm{probe}}$ & Number of probe pixels \\
\hline
$L$ & Number of optimisation iterations (epochs) \\
\hline
$M$ & Number of probe modes in the reconstruction \\
\hline
$F$ & Number of diffraction patterns per \texttt{PtyPy} data block \\
\hline
\end{tabular}
\end{table}

The memory footprint of the VR algorithm is dominated by the diffraction data held on the device and by the object together with its optimiser state. In the configuration benchmarked here all $K$ measured patterns are held in GPU memory, so the resident data scales with the size of the dataset. The mini-batch size enters only through the tensors retained for reverse-mode differentiation, not through the resident data. The peak memory, including data-type dependent constants, is

\begin{equation}
\label{VR memory equation}
\begin{aligned}
    mem_{VR} \approx{}&
    mem_{\mathrm{torch}}
    + c_{\mathrm{NAdam}}\,c_{\mathrm{complex64}}\,N_{\mathrm{params}}
    + c_{\mathrm{float32}}\,K N_{\mathrm{probe}}
    + c_{\mathrm{BK}} K \\
    &+ \max\!\left( c_{\mathrm{act}}(M)\, K' N_{\mathrm{probe}},\;
                    c_{\mathrm{opt}}\, (N_{\mathrm{object}} + M N_{\mathrm{probe}} + K M) \right) .
\end{aligned}
\end{equation}

Here, $mem_{\mathrm{torch}}$ is the fixed overhead of the CUDA/PyTorch runtime, measured as $0.65$~GB from the first sample of the memory monitor before any data is loaded. The object, the probe modes and the $K$ per-pattern scaling factors $\kappa_k$ of Eq.~\ref{Scaling and Incoherent Modes} are stored as single-precision complex numbers, $c_{\mathrm{complex64}} = 8~\text{bytes}$, and NAdam holds four arrays of this size for each of them: the parameter, its gradient and two moment estimates, so that $c_{\mathrm{NAdam}} = 4$. Diffraction data is stored as single-precision reals, $c_{\mathrm{float32}} = 4~\text{bytes}$, and $c_{\mathrm{BK}} = 28~\text{bytes}$ accounts for per-pattern bookkeeping held on the device: the sampling weight of Eq.~\ref{Sampling Bias} (4~bytes), the index vector drawn for each epoch (8~bytes) and the integer scan position (16~bytes).

\begin{table}[H]
\centering
\caption{Transient memory of one VR reconstruction step, measured from the PyTorch allocator as the slope against $K' N_{\mathrm{probe}}$ over a sweep of $K'$ sizes, with the data, parameters and optimiser moments already allocated.}
\label{tab:activation-constants}
\begin{tabular}{|c|c|c|c|c|}
\hline
\textbf{Probe modes $M$} & 1 & 2 & 3 & 4 \\ \hline
$c_{\mathrm{act}}(M)$ (bytes) & 60 & 80 & 112 & 144 \\ \hline
\end{tabular}
\end{table}

The term $\max\!\left( c_{\mathrm{act}}(M)\, K' N_{\mathrm{probe}},\; c_{\mathrm{opt}}\, (N_{\mathrm{object}} + M N_{\mathrm{probe}} + K M) \right)$ is the transient memory of one reconstruction step. The forward propagation of a mini-batch and its gradient allocate $c_{\mathrm{act}}(M)$ bytes per detector pixel of the mini-batch; the optimiser step then allocates a temporary copy of every parameter, $c_{\mathrm{opt}} = 8$~bytes per element, after the mini-batch tensors have been released. $c_{\mathrm{act}}(M)$ is measured from the maximum GPU memory allocated by the PyTorch allocator during one update step, as the slope against $K' N_{\mathrm{probe}}$ over a sweep of $K'$ (Tab.~\ref{tab:activation-constants}). It is $60$~bytes for a single mode, where the peak falls in the evaluation of the Poisson negative log-likelihood, and $16 + 32M$ bytes for $M \geq 2$, where the peak falls in the backward pass through the exit waves of every mode. Eq.~\ref{VR memory equation} counts allocated tensors; the per-process memory reported by the \textit{nvidia-smi} driver, which Section~\ref{sec:Scaling results} measures, also includes what the caching allocator holds in reserve, and so exceeds the count predicted by these scaling equations by a small margin.

The computational cost of the VR algorithm is dominated by FFT operations in the forward model and its adjoint, with a second contribution from the optimiser, which updates every parameter once per mini-batch step. We write the wall-clock time to convergence as $T_{VR} = L\, t_{\mathrm{epoch}}$, with

\begin{equation}
\label{VR time equation}
    t_{\mathrm{epoch}} \approx
      a_{\mathrm{VR}}\, K M N_{\mathrm{probe}}
    + \left\lceil \frac{K}{K'} \right\rceil
      \left( b_{\mathrm{VR}}\, N_{\mathrm{params}} + g_{\mathrm{VR}} \right) .
\end{equation}

Although mini-batches of size $K'$ are used for individual updates, we consider one iteration of the algorithm to be one epoch of the mini-batch method. One epoch is a single pass through all $K$ patterns, so a mini-batch method with batch size $K'$ performs $\left\lceil \frac{K}{K'} \right\rceil$ parameter-update steps per epoch, with the final mini-batch containing $K - K' \left\lfloor \frac{K}{K'} \right\rfloor$ patterns when $K$ is not an integer multiple of $K'$. Typically, multiple epochs are run so that the full dataset is revisited several times, and reporting progress in epochs rather than raw iterations is standard in stochastic-gradient training and gives a more comparable measure across methods with different mini-batch sizes~\cite{Bottou_2016}.

The first term is the cost of propagating $K$ patterns through $M$ modes. The constant $a_{\mathrm{VR}}$ is the time per detector pixel per mode per pass and absorbs the forward and inverse Fourier transforms of the exit waves, the element-wise operations of the forward model and the likelihood, and their adjoints. $a_{\mathrm{VR}}$ is measured as the slope of the step time against $K' M N_{\mathrm{probe}}$ over a sweep of the mini-batch size at fixed object, along which the other two terms of Eq.~\ref{VR time equation} are constant (Tab.~\ref{tab:time-constants}).

The second term accounts for the optimiser. NAdam updates the whole parameter set at every mini-batch step, and an epoch contains $\lceil K/K' \rceil$ such steps, so the optimiser cost per epoch grows as the mini-batch size is decreased. $b_{\mathrm{VR}}$ is measured as the slope of the step time against $N_{\mathrm{params}}$ over a sweep of the object size at fixed mini-batch.

\paragraph{Computational Complexity of the Maximum-Likelihood Algorithm}
As a baseline, we consider the maximum-likelihood (ML) algorithm implemented in \texttt{PtyPy}, which employs full-batch first-order gradient updates under a Gaussian statistical noise model. \texttt{PtyPy} subdivides the diffraction data into blocks of $F$ patterns and manages device residency block by block, so the dominant memory contributions are the resident blocks and the working buffers allocated for a single block. The peak memory, including constant factors, is

\begin{equation}
\label{ML memory equation}
\begin{aligned}
    mem_{ML} \approx{}&
    mem_{CuPy}
    + c_{\mathrm{CG}}\,c_{\mathrm{complex64}}
      \left(N_{\mathrm{object}} + M N_{\mathrm{probe}}\right) \\
    &+ 2 c_{\mathrm{float32}}
      \left\lceil \frac{K}{F} \right\rceil F N_{\mathrm{probe}}
    + \left(3 c_{\mathrm{complex64}} M + 3 c_{\mathrm{float32}}\right)
      F N_{\mathrm{probe}} .
\end{aligned}
\end{equation}

Here, $c_{\mathrm{CG}} = 4$ accounts for storage of the object and probe together with the gradient, the trial gradient and the conjugate search direction. Both algorithms are run in single precision, so the same data-type constants $c_{\mathrm{complex64}} = 8~\text{bytes}$ and $c_{\mathrm{float32}} = 4~\text{bytes}$ apply, giving $mem_{ML}$ as an estimate of memory usage in bytes. The factor of two on the resident data accounts for the measured intensities and the per-pattern weights, both stored as single-precision reals. As only whole blocks are allocated, the resident data rounds up to a block boundary and exceeds $2 c_{\mathrm{float32}} K N_{\mathrm{probe}}$ by up to one block. The final term is the working set of a single block: three complex buffers used by the propagation and gradient kernels, and three real buffers holding the modelled intensity and the likelihood accumulators. It is independent of $K$, and for the smallest dataset of Section~\ref{sec:Scaling results} it is the largest single contribution to the footprint. $mem_{CuPy}$ is the memory overhead associated with the CuPy context, measured as $0.55$~GiB.

The time of one ML iteration is likewise dominated by the propagation of every pattern, here performed three times: for the gradient, for the trial step of the polynomial line search and for the update. We write $T_{ML} = L\, t_{\mathrm{iter}}$, with

\begin{equation}
\label{ML time equation}
    t_{\mathrm{iter}} \approx
      a_{\mathrm{ML}}\, K M N_{\mathrm{probe}}
      + g_{\mathrm{ML}} .
\end{equation}

The constant $a_{\mathrm{ML}}$ is the time per detector pixel per mode per iteration, which was measured as the slope of the iteration time against $K M N_{\mathrm{probe}}$ over seven synthetic datasets of $1.5 \times 10^{3}$ to $2.5 \times 10^{5}$ patterns, with the parameter tree and block size of the reconstructions of Section~\ref{sec:Scaling results}.

The ML algorithm performs a single full-batch update per iteration, so the latency $g_{\mathrm{ML}}$ is incurred once per iteration rather than once per diffraction pattern; it is the intercept of the same sweep and is consistent with zero.

\begin{table}[H]
\centering
\caption{Time constants of Eqs.~\ref{VR time equation} and~\ref{ML time equation}, measured by benchmarks on the NVIDIA H100 used for every reconstruction, for $128 \times 128$ diffraction patterns and a single probe mode. Each constant is the slope or intercept of one sweep in which only the corresponding quantity changes.}
\label{tab:time-constants}
\begin{tabular}{|c|c|c|p{0.35\textwidth}|}
\hline
\textbf{Constant} & \textbf{Value} & \textbf{Std.\ error} & \textbf{Unit} \\ \hline
$a_{\mathrm{VR}}$ & $2.48 \times 10^{-10}$ & $9.0 \times 10^{-13}$ & s per $K' M N_{\mathrm{probe}}$ element \\ \hline
$b_{\mathrm{VR}}$ & $5.84 \times 10^{-11}$ & $9.7 \times 10^{-13}$ & s per parameter element \\ \hline
$g_{\mathrm{VR}}$ & $4.45 \times 10^{-4}$ & $4.9 \times 10^{-5}$ & s per step \\ \hline
$a_{\mathrm{ML}}$ & $2.85 \times 10^{-10}$ & $2.4 \times 10^{-12}$ & s per $K M N_{\mathrm{probe}}$ element \\ \hline
$g_{\mathrm{ML}}$ & $3.37 \times 10^{-3}$ & $4.3 \times 10^{-3}$ & s per iteration \\ \hline
\end{tabular}
\end{table}

\paragraph{Comparison of Computational Complexities}
The VR algorithm operates on mini-batches of size $K'$, whereas the ML algorithm evaluates gradients using the full dataset of size $K$ at every update. Comparing Eq.~\ref{VR memory equation} with Eq.~\ref{ML memory equation}, the memory advantage of VR arises from three distinct sources. First, VR stores only the measured intensities, whereas \texttt{PtyPy} stores intensities and per-pattern noise weights to calculate the Gaussian negative log likelihood model, a factor of two in bytes per detector pixel. Second, \texttt{PtyPy} allocates whole blocks, so its resident data rounds up to a multiple of $F$. Third, the transient working set of VR scales with the mini-batch size $K'$ while that of \texttt{PtyPy} scales with the block size $F$.

The ML memory expression above assumes that every block is resident on the device. \texttt{PtyPy} relaxes this by capping the number of resident blocks at what fits in free memory and streaming the remainder between host and device, so that full-batch ML can run on hardware that cannot hold the entire dataset. Data blocking does not alter the asymptotic comparison with VR, as the number of patterns processed per epoch, and hence the FFT count and floating-point cost, is unchanged. It does, however, introduce repeated host-to-device transfers whose bandwidth and latency overhead scales with $K$ and is not captured by $g_{\mathrm{ML}}$.

The time per iteration of both methods is dominated by the FFT in the ptychography model model and therefore exhibits the same dependence on $K$, $M$ and $N_{\mathrm{probe}}$. The per-pixel time constants, measured per mode for one ML iteration and one VR epoch respectively, are similar: $a_{\mathrm{ML}}$ is approximately $15\%$ larger than $a_{\mathrm{VR}}$ (Tab.~\ref{tab:time-constants}). The effective computational cost per parameter update differs by a factor of approximately $K/K'$, because ML processes all diffraction patterns at each update whereas VR processes only a subset. On an epoch basis, however, VR is not the cheaper method: it performs a comparable number of propagations per pass over the data, and additionally sweeps the parameter set $\lceil K/K' \rceil$ times.

\subsection{Similar Work}

Several recent works address related challenges in ptychographic reconstruction, particularly in computational efficiency, scalability, and modelling flexibility. We briefly review prominent examples in the remainder of this section.

Wu \textit{et al.} propose an automatic-differentiation framework for low-dose ptychography (DAP)~\cite{Wu_2024}. DAP minimises a Poisson or Gaussian negative log-likelihood with Adam and reduces gradient noise by increasing the mini-batch size over the course of the reconstruction alongside a plateau-based learning-rate schedule. It is evaluated on simulated data and on measured datasets of up to $4\times10^{4}$ diffraction patterns against ePIE and ADMM. The present work differs in three respects. First, we hold the mini-batch size fixed, so peak memory is set by $K'$ rather than growing during the reconstruction. Second, wall-clock time and peak GPU memory are measured against the full-batch ML algorithm of \texttt{PtyPy} and are predicted by the time and memory models of Section~\ref{subsec:Computational_Complexity_Analysis}. Third, the evaluation extends to $2\times10^{5}$ experimental and $1.02\times10^{6}$ synthetic diffraction patterns.

Nashed \textit{et al.} introduce a multi-GPU framework for ptychography~\cite{Nashed_2017} and study its scaling behaviour across multiple GPUs using synthetic data. Our work differs from the multi-GPU framework of Nashed \textit{et al.} in several key aspects: we directly compare against a state-of-the-art ptychography package (\texttt{PtyPy}) on experimental datasets; we demonstrate performance down to a single GPU (rather than a minimum of four GPUs for synthetic and 32 GPUs for experimental data in~\cite{Nashed_2017}); and we evaluate a range of experimental conditions with varying memory footprints, frame counts, and probe/object sizes. Furthermore, Nashed \textit{et al.} use in their simulation a synthetic dataset of $2.56 \times 10^4$ diffraction patterns, $128 \times 128$ pixels per diffraction pattern. This is an order of magnitude smaller than the largest synthetic dataset which we consider in Section~\ref{sec:Scaling results} of size $128 \times 128$ pixels per diffraction pattern, $1.02 \times 10^6$ diffraction patterns. Additionally, the experimental dataset used by Nashed \textit{et al.} of $4.2 \times 10^5$ diffraction patterns of size $128 \times 128$ pixels is smaller than the largest experimental dataset considered in this work as detailed in Table~\ref{Experimental Datasets Table}. In addition, we explicitly model probe instability.

Ghosh \textit{et al.} propose an automatic-differentiation-based ptychographic reconstruction framework~\cite{Ghosh_2018}. The automatic-differentiation framework of Ghosh \textit{et al.} does not consider variance-reduced optimisation, nor does it analyse scaling with problem size or evaluate performance across diverse experimental conditions. Our method, by contrast, combines a differentiable forward model with variance reduction and an explicit Poisson noise model, and includes multi-modal probe modelling to account for partial coherence and probe fluctuations.

Du \textit{et al.} present Adorym, a general framework for x-ray imaging reconstruction~\cite{Du_2020}. While it supports flexible forward models, it does not incorporate variance reduction and relies primarily on a single experimental dataset for 2D ptychography, without systematic comparison to other reconstruction algorithms. Our work focuses specifically on variance-reduced optimisation in ptychography and provides direct comparisons to established methods.

Lee \textit{et al.} introduce PtyRAD, which integrates automatic differentiation into electron ptychography reconstruction~\cite{Lee_2025}. The PtyRAD reconstruction approach similarly does not incorporate dynamic variance reduction. In contrast, our method applies dynamic variance reduction to x-ray ptychography, analyses computational complexity, and is benchmarked across multiple experimental and synthetic datasets.

\section{Experimental Setup}
\label{sec:Experimental_Setup}

In this section, we evaluate the proposed variance-reduction (VR) reconstruction algorithm in terms of reconstruction fidelity, computational efficiency, and scaling behaviour.

Reconstruction fidelity is assessed on experimental datasets using the Gaussian negative log-likelihood described in Section~\ref{Loss Function Details}, together with visual inspection of object reconstructions. The Gaussian negative log-likelihood serves as the common metric across all algorithms: for ePIE, DM, and ML it is obtained from the per-iteration error values recorded by the PtyPy package. ML minimises the Gaussian NLL, while DM and ePIE are set-projection algorithms that enforce a Fourier-magnitude constraint rather than minimising a statistical likelihood. VR minimises the Poisson negative log-likelihood of Eq.~\ref{PoissonNLL Equation}, rather than the Gaussian NLL used here as a common reporting metric. A higher Gaussian NLL for VR therefore does not necessarily indicate poorer optimisation, but reflects the differing objectives of the reconstruction methods; correspondingly, VR may attain a lower Poisson NLL than other algorithms. Computational efficiency is quantified by the cumulative runtime required to satisfy the convergence criteria defined in Sec.~\ref{Convergence Criteria}, and by peak GPU memory consumption during the execution of the reconstruction script. We report time to convergence as the cumulative iteration time, excluding data loading, up to the point at which the convergence criterion is satisfied. Peak GPU memory usage is measured over the complete execution of the program, including data loading.

We compare the VR algorithm against Difference Map (DM), ePIE, and Maximum Likelihood (ML) implementations from PtyPy. Particular emphasis is placed on the comparison between ML and VR, because both reconstruction algorithms are formulated using statistical noise models. GPU memory usage is a central focus because reducing memory complexity is a principal motivation for introducing stochastic variance-reduced optimisation in large-scale ptychographic reconstruction.

The VR reconstruction routine is implemented in Python using PyTorch~\cite{Paszke_2019}, with data handled via h5py~\cite{Collette_2013}. The reconstruction is formulated as the minimisation of the Poisson negative log-likelihood, as defined in Eq.~\ref{PoissonNLL Equation}, which is the sole optimisation objective used by VR. The sampling distribution for gradient updates is defined as a function of the power of each diffraction pattern, as given in Equation~\ref{Sampling Bias}. Initialising the object with unit pixel-wise magnitude and small random phase ensures consistent gradient magnitudes across datasets and matches the default strategy used in PtyPy. The minimiser is approximated using the NAdam optimiser, implementing a variant of stochastic gradient descent with adaptive updates. Comparisons are performed against PtyPy~\cite{Ptypy}.

All reconstructions were performed on a Linux-based high-performance computing cluster equipped with a 48-core CPU, 1.5~TB of system memory, and a single NVIDIA H100 GPU with 94~GB of device memory. The VR mini-batch size $K'$ was chosen for each dataset as a power of two, between $1.8\%$ and $5.9\%$ of the number of diffraction patterns for the experimental datasets; the values used are listed in Section~\ref{sec:Experimental_Demonstration}. These choices were guided by practical considerations and were found empirically to yield stable convergence behaviour.

\subsection{Convergence Criteria}
\label{Convergence Criteria}

Let $\mathcal{L}_P$ denote the Poisson negative log-likelihood (NLL), and let $\{\mathcal{L}_k\}_{k=0}^{N-1}$ denote the sequence of loss values evaluated at each iteration (or epoch). Convergence is assessed by monitoring the trend of a smoothed loss curve, calculated using the loss curve reported by each algorithm. Smoothing reduces the influence of small fluctuations around the loss function minimum, thereby providing a more robust estimate of the point of convergence. We refer to this rule as the loss-plateau criterion. It tests the slope of the smoothed loss with respect to the iteration number, not the gradient of the objective with respect to the reconstruction parameters.

The loss sequence is first smoothed using an exponential moving average (EMA),
\begin{equation}
S_k = \alpha \mathcal{L}_k + (1-\alpha)S_{k-1},
\end{equation}
where $S_k$ denotes the smoothed loss, $S_0=\mathcal{L}_0$, and $\alpha=0.095$. This value corresponds to an effective averaging window of approximately twenty epochs.

For each epoch $k \geq 19$, a sliding window comprising of the twenty most recent smoothed loss values is defined as
\begin{equation}
W_k = {S_{k-19}, \ldots, S_k}.
\end{equation}

The local trend of the smoothed loss within each window is quantified by fitting a first-order polynomial to the values in $W_k$ using ordinary least-squares regression. Let $m_k$ denote the resulting slope coefficient, which represents the average rate of change of the smoothed loss over the preceding twenty epochs. The convergence metric is then defined as
\begin{equation}
C_k = |m_k|.
\end{equation}

To account for scale variations in the loss function, a normalized slope metric is also computed,
\begin{equation}
\tilde{C}_k =
\frac{C_k}
{\left|
\frac{1}{20}\sum_{j=k-19}^{k} S_j
\right|
+ \varepsilon}.
\end{equation}
where $\varepsilon = 10^{-12}$ ensures numerical stability.

Convergence is declared at the first iteration $k^\ast$ such that

\begin{equation}
\label{eq:Gradient_Stopping_Condition}
    C_k < \tau_{\mathrm{abs}}
    \quad \text{or} \quad
    \tilde{C}_k < \tau_{\mathrm{rel}},
\end{equation}

with thresholds $\tau_{\mathrm{abs}} = 10^{-4}$ and $\tau_{\mathrm{rel}} = 10^{-5}$. The threshold values were selected empirically based on preliminary experiments spanning a representative range of datasets and initial conditions. The final values were chosen to provide stable convergence detection across all tested cases. If no such $k^\ast$ is identified, the algorithm proceeds until reaching the maximum number of pre-specified iterations. In all experiments, the pre-determined maximum number of epochs was chosen to be sufficiently large such that the loss-plateau criterion of Equation~\ref{eq:Gradient_Stopping_Condition} was reached before termination.

\section{Experimental Demonstration}
\label{sec:Experimental_Demonstration}

We evaluate reconstruction fidelity and computational efficiency on three experimental datasets acquired under varying imaging conditions; see Tab.~\ref{Experimental Datasets Table}. The three experimental datasets span photon-counting and non-photon-counting detector types, as well as a wide range of diffraction-frame counts and object field of views.

\paragraph{Experimental Datasets}
To assess reconstruction quality and computational performance, three experimental datasets were used, which span a variety of experimental conditions. All datasets were collected at the Diamond Light Source. The first dataset are 200,000 ptychographic diffraction patterns from a tungsten test target collected at the I14 nanoprobe instrument with the photon-counting detector (Merlin Quad, Quantum Detectors, UK) at a photon energy of 12 keV. The second dataset are 900 ptychographic diffraction patterns from a sample with Au nanoparticles deposited on a silicon-nitride membrane collected at the I08-1 instrument with the charge-integrating sCMOS detector (AXIS-SXRF-2020EUV, Axis Photonique Inc.) at a photon energy of 710 eV. The third dataset are 17,496 ptychographic diffraction patterns from a brown cover scale taken from a butterfly of the Lycaenid family collected at the I08-1 instrument with the sCMOS Axis Photonique detector at a photon energy of 1 keV. Furthermore, a smaller butterfly scale dataset is used which is cropped to a subset of 2,346 diffraction pattern. In addition to the collection of diffraction patterns, each dataset also includes corresponding probe positions, and other experimental parameters that are required for ptychographic reconstructions. A full list of all the datasets is provided in  in Tab.~\ref{Experimental Datasets Table} including the size of the reconstructed objects in pixel dimensions.

\begin{table}[H]
\centering
\caption{Experimental datasets used for algorithm validation.}
\label{Experimental Datasets Table}
\begin{tabular}{|c|c|c|c|}
\hline
\textbf{Dataset} & 
\shortstack{\textbf{Number of}\\\textbf{diffraction patterns}} & 
\shortstack{\textbf{Diffraction pattern}\\\textbf{pixel dimensions}} & 
\shortstack{\textbf{Object size}\\\textbf{pixel dimensions}} \\
\hline
Nanoprobe Test Target & 200{,}000 & $128 \times 128$ & $1578 \times 1941$ \\
\hline
Au Particles & 900 & $512 \times 512$ & $1095 \times 1103$ \\
\hline
Butterfly Scale (large) & 17{,}496 & $512 \times 512$ & $6233 \times 2842$ \\
\hline
Butterfly Scale (small) & 2346 & $512 \times 512$ & $1638 \times 1806$ \\
\hline
\end{tabular}
\end{table}

All diffraction data were stored as 64-bit integer arrays. Reconstructions using the VR algorithm were carried out in \texttt{PyTorch} using complex64 tensors.

\paragraph{Reconstruction hyperparameters and model parameters} The Au particle dataset was reconstructed with the probe modelled using three modes to ensure convergence, for both ML and VR. The butterfly scale large and butterfly scale small datasets were reconstructed using two probe modes to ensure convergence, for all reconstruction algorithms. Apart from the number of probe modes, which was the same for both algorithms, two VR settings were chosen for each dataset: the mini-batch size $K'$ and the \textit{factor} parameter. All other VR hyperparameters were identical across all experimental and synthetic datasets. For ML, three settings were chosen for each experimental dataset: the block size $F$, and whether the \texttt{PtyPy} scale preconditioner and floating intensities were enabled. The \textit{factor} parameter controls the proportional reduction in the update step size each time the learning-rate scheduler is applied. The \texttt{PtyPy} block sizes were $F = 3000$, $900$, $300$ and $2500$ patterns for the nanoprobe test target, Au particles, butterfly scale (large) and butterfly scale (small) datasets respectively. $F = 6000$ for every synthetic dataset. The corresponding VR mini-batch sizes were $K' = 4096$, $16$, $1024$ and $64$. Neither $F$ nor $K'$ was tuned to the GPU used, so the reported memory and runtimes reflect these configurations rather than the minimum achievable by either algorithm.

\begin{figure}[H]
    \centering
    \includegraphics[clip, width=\textwidth]{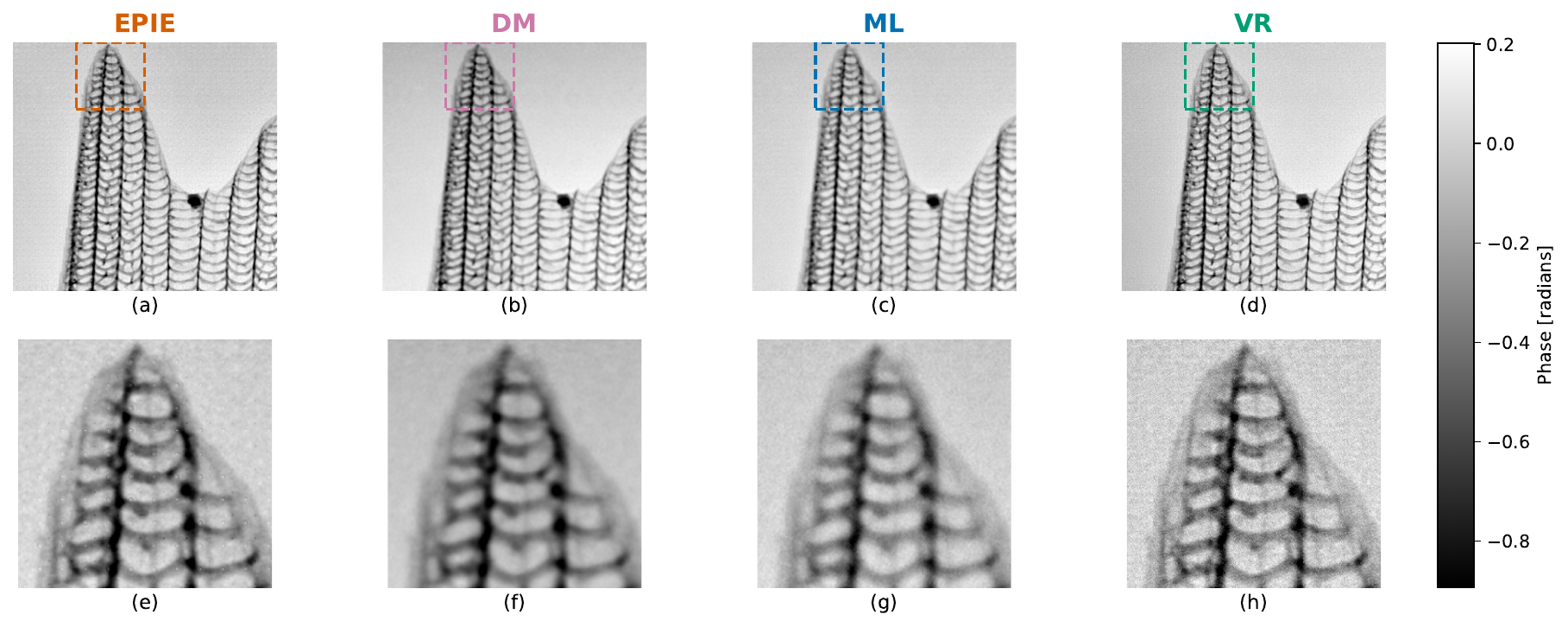}
    \caption{
Phase reconstructions of the butterfly scale small specimen. The top row shows the full-field reconstructions obtained using (a) ePIE, (b) DM, (c) ML, and (d) VR. The bottom row shows the corresponding $5\,\mu\mathrm{m} \times 5\,\mu\mathrm{m}$ magnified region for (e) ePIE, (f) DM, (g) ML, and (h) VR. All panels share a common phase colour scale given in radians. Dashed squares indicate the region shown at higher magnification in the bottom row.}
    \label{Cropped butterfly object reconstruction comparison}
\end{figure}

\paragraph{Reconstruction Results}
Figure~\ref{Cropped butterfly object reconstruction comparison} compares object phase reconstructions for the butterfly scale small dataset. The top row, panels (a)–(d), shows reconstructions at termination according to the loss-plateau criterion, while the bottom row, panels (e)–(h), shows the corresponding $5\,\mu\mathrm{m} \times 5\,\mu\mathrm{m}$ subregion indicated by the dashed boxes.

At the stopping point of the loss-plateau criterion, all four algorithms recover the structural features of the object, but differ in sharpness, artefact presence, and noise levels. The ePIE (Fig.~\ref{Cropped butterfly object reconstruction comparison}(a)) and VR (Fig.~\ref{Cropped butterfly object reconstruction comparison}(d)) reconstructions exhibit comparable overall sharpness, whereas the DM (Fig.~\ref{Cropped butterfly object reconstruction comparison}(b)) and ML (Fig.~\ref{Cropped butterfly object reconstruction comparison}(c)) reconstructions appear comparatively blurred. These differences are more apparent in the magnified regions: ePIE (Fig.~\ref{Cropped butterfly object reconstruction comparison}(e)) exhibits repeated spot-like artefacts, whilst ML (Fig.~\ref{Cropped butterfly object reconstruction comparison}(g)) retains greater fine-scale detail than DM (Fig.~\ref{Cropped butterfly object reconstruction comparison}(f)) but with slightly increased noise. VR (Fig.~\ref{Cropped butterfly object reconstruction comparison}(h)) preserves the greatest level of fine-scale detail in the magnified region, although this is accompanied by a higher apparent noise level. Thus, at this stopping point, the reconstructions exhibit a trade-off between preservation of fine-scale structure and noise, with VR retaining the most detail while DM produces the smoothest appearance.

Fig.~\ref{Cropped butterfly object reconstruction comparison} hence demonstrates that under the aforementioned convergence criterion, the VR algorithm can attain results which are consistent with those attainable by existing ptychography reconstruction algorithms.

\begin{figure}[H]
    \centering
    \includegraphics[clip, width=0.7\textwidth]{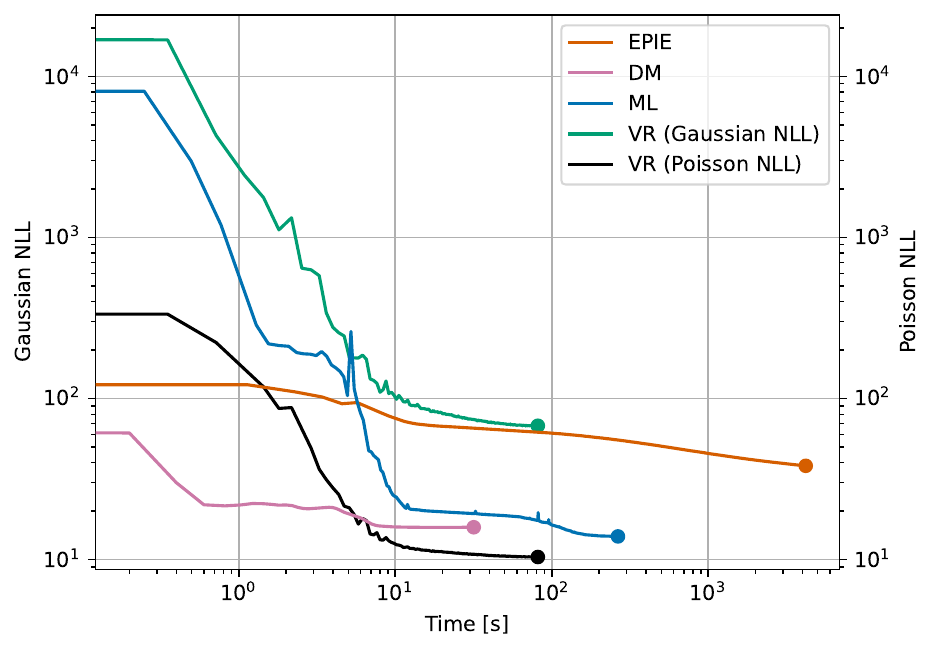}
    \caption{Gaussian log-likelihood loss as a function of cumulative runtime for reconstructions of the butterfly scale small dataset using DM, ML, ePIE, and VR. Both axes are shown on logarithmic scales. Solid lines denote iterations up to the stopping point of the loss-plateau criterion, with circular markers indicating the stopping point. VR minimises the Poisson NLL loss, ML minimises the Gaussian NLL, and DM and ePIE minimise a Fourier-magnitude constraint error.}
    \label{Cropped butterfly loss curve comparison}
\end{figure}

Figure~\ref{Cropped butterfly loss curve comparison} shows the Gaussian log-likelihood loss curves corresponding to the reconstructions in Fig.~\ref{Cropped butterfly object reconstruction comparison}. The ML algorithm achieves the lowest Gaussian negative log-likelihood loss of all reconstruction algorithms considered. The four algorithms minimise different objectives---VR the Poisson NLL, ML the Gaussian NLL, and DM and ePIE a Fourier-magnitude constraint error---so their loss values reflect these differing formulations as well as the quality of the reconstructed object.

Each algorithm is assessed individually and is run from the same generic initialisation; no algorithm is warm-started from the reconstruction produced by another, and in particular VR is not initialised from a DM or ePIE reconstruction. This is a deliberate choice: applying VR directly to the measured data, without a prior full-batch reconstruction, is what preserves its low peak-memory footprint, and computing such an initialisation would reintroduce the memory cost that VR is designed to avoid. The lower initial loss of DM and ePIE in Fig.~\ref{Cropped butterfly loss curve comparison} is likewise a property of the algorithm class rather than of the starting point: set-projection algorithms apply an exact correction to each constraint set at every iteration, whereas the gradient-based methods minimise a loss function under a step-size schedule chosen to promote stable convergence.

The loss curves of Fig.~\ref{Cropped butterfly loss curve comparison} are indicative of convergence behaviour; the absolute ``converged'' loss values are only a partial indicator of the quality of the object reconstruction. The reconstructions shown in Fig.~\ref{Cropped butterfly object reconstruction comparison} indicate that it is not particularly informative to attempt to delineate between converged loss values within the same order of magnitude.

Tab.~\ref{tab:cropped-butterfly-reconstruction-statistics} summarises the quantitative performance of each algorithm on the butterfly scale small dataset.

\newcommand{\EPIELoss}{38.22}
\newcommand{\EPIETime}{4195.89}
\newcommand{\EPIEMemory}{5.53}

\newcommand{\DMLoss}{15.88}
\newcommand{\DMTime}{31.73}
\newcommand{\DMMemory}{30.18}

\newcommand{\MLLoss}{13.93}
\newcommand{\MLTime}{264.78}
\newcommand{\MLMemory}{47.45}

\newcommand{\VRLoss}{67.78}
\newcommand{\VRTime}{81.50}
\newcommand{\VRMemory}{5.20}

\begin{table}[H]
\centering
\caption{Reconstruction statistics for the cropped butterfly dataset. The terminal loss is evaluated at the point of convergence defined by the loss-plateau criterion. The loss is the Gaussian negative log-likelihood recorded by PtyPy; VR, which optimises the Poisson negative log-likelihood, records it in a separate, untimed run. Peak GPU memory is the memory attributed to the reconstruction process, over a window that spans data loading; the reported time covers the iterations alone. Experiments were run on an NVIDIA H100 with 94 GB of device memory.}
\label{tab:cropped-butterfly-reconstruction-statistics}
\begin{tabular}{|c|c|c|c|}
\hline
\textbf{Algorithm} & \shortstack{\textbf{Gaussian Negative} \\ \textbf{Log-Likelihood Loss}} & \textbf{Time to convergence (s)} & \shortstack{\textbf{Peak GPU} \\ \textbf{memory usage (GB)}}\\
\hline
ePIE & \EPIELoss & \EPIETime & \EPIEMemory \\
\hline
DM & \DMLoss & \DMTime & \DMMemory \\
\hline
ML & \MLLoss & \MLTime & \MLMemory \\
\hline
VR & \VRLoss & \VRTime & \VRMemory \\
\hline
\end{tabular}
\end{table}

The DM algorithm satisfies the loss-plateau criterion in the lowest cumulative runtime, approximately twice as fast as the VR algorithm. In practical workflows however, DM is typically employed as an initialisation strategy for ML refinement~\cite{Scala_2018}. When this subsequent optimisation stage is accounted for, the effective time to convergence increases substantially, reducing the apparent computational advantage of DM when considered in isolation. VR, by contrast, is applied directly to the raw data with no initialising reconstruction, so its reported runtime and memory figures reflect the complete cost of the method.

Peak GPU memory consumption is reduced from \MLMemory \ GB (ML) to \VRMemory  \ GB (VR), a factor of nine relative to ML and of six relative to DM. These factors reflect the configurations compared as well as the algorithms: with a block size of $F = 2500$, \texttt{PtyPy} holds all $2346$ diffraction patterns in a single block, and Eq.~\ref{ML memory equation} predicts approximately $6.6$~GB for ML with $F$ equal to the VR mini-batch size of $K' = 64$. ePIE, which updates the object one diffraction pattern at a time, has a footprint comparable to that of VR (\EPIEMemory \ GB) but takes fifty times longer to satisfy the convergence criterion.

Taken together, the reconstruction-quality, loss, runtime, and memory measurements indicate that VR achieves object reconstructions of comparable visual fidelity whilst maintaining close to an order-of-magnitude advantage in GPU memory usage over the full-batch algorithms at the block size used, and a comparatively quick reconstruction time.

Tab.~\ref{tab:experimental-dataset-reconstruction-statistics-ml-vr} reports reconstruction statistics for the full experimental datasets using the ML and VR algorithms. Peak GPU memory usage is lower for VR on every dataset, by factors of between $1.5{\times}$ and $10.8{\times}$. The $10.8{\times}$ memory saving on the Au particles dataset arises because the \texttt{PtyPy} block size of $F = 900$ equals the number of diffraction patterns. Consequently, a single block holds the entire dataset and memory allocation for ML scales with $F$ and the three probe modes, whereas the VR working set scales with its mini-batch size of $K' = 16$ (Eq.~\ref{ML memory equation}).

\newcommand{\IFOURTEENLogoMLLoss}{5.90}
\newcommand{\IFOURTEENLogoMLTime}{744.23}
\newcommand{\IFOURTEENLogoMLMemory}{28.87}
\newcommand{\IFOURTEENLogoVRLoss}{19.76}
\newcommand{\IFOURTEENLogoVRTime}{165.29}
\newcommand{\IFOURTEENLogoVRMemory}{18.76}
\newcommand{\NanogoldMLLoss}{5.63}
\newcommand{\NanogoldMLTime}{770.18}
\newcommand{\NanogoldMLMemory}{26.13}
\newcommand{\NanogoldVRLoss}{10.07}
\newcommand{\NanogoldVRTime}{76.82}
\newcommand{\NanogoldVRMemory}{2.42}
\newcommand{\ButterflyMLLoss}{15.82}
\newcommand{\ButterflyMLTime}{1328.82}
\newcommand{\ButterflyMLMemory}{80.12}
\newcommand{\ButterflyVRLoss}{83.97}
\newcommand{\ButterflyVRTime}{563.62}
\newcommand{\ButterflyVRMemory}{50.54}

\begin{table}[H]
\centering
\caption{Reconstruction statistics for full experimental datasets comparing ML and VR. The terminal loss is evaluated at the point of convergence defined by the loss-plateau criterion. The loss is the Gaussian negative log-likelihood recorded by PtyPy; VR, which optimises the Poisson negative log-likelihood, records it in a separate, untimed run. Peak GPU memory is the memory attributed to the reconstruction process, over a window that spans data loading. Experiments were run on an NVIDIA H100 with 94 GB of device memory.}
\label{tab:experimental-dataset-reconstruction-statistics-ml-vr}
\begin{tabular}{|c|c|c|c|c|c|c|}
\hline
\multirow{2}{*}{\textbf{Dataset}} & \multicolumn{2}{c|}{\shortstack{\textbf{Gaussian Negative} \\ \textbf{Log-Likelihood Loss}}} & \multicolumn{2}{c|}{\textbf{Time to convergence (s)}} & \multicolumn{2}{c|}{\shortstack{\textbf{Peak GPU} \\ \textbf{memory usage (GB)}}} \\
\cline{2-7}
 & ML & VR & ML & VR & ML & VR \\
\hline
Nanoprobe test target & \IFOURTEENLogoMLLoss & \IFOURTEENLogoVRLoss & \IFOURTEENLogoMLTime & \IFOURTEENLogoVRTime & \IFOURTEENLogoMLMemory & \IFOURTEENLogoVRMemory \\
\hline
Au particles & \NanogoldMLLoss & \NanogoldVRLoss & \NanogoldMLTime & \NanogoldVRTime & \NanogoldMLMemory & \NanogoldVRMemory \\
\hline
Butterfly scale (large) & \ButterflyMLLoss & \ButterflyVRLoss & \ButterflyMLTime & \ButterflyVRTime & \ButterflyMLMemory & \ButterflyVRMemory \\
\hline
\end{tabular}
\end{table}

The summary statistics of Tab.~\ref{tab:experimental-dataset-reconstruction-statistics-ml-vr} give similar indications of the relative performance of the VR algorithm relative to the ML algorithm as demonstrated on the butterfly scale small dataset, now demonstrated on larger datasets. VR reduces the reconstruction time and, at the block sizes used, the peak memory usage, which allows VR to be applied to larger problem sizes where memory is a limiting constraint.

\begin{figure}[H]
    \centering
    \includegraphics[clip, width=\textwidth]{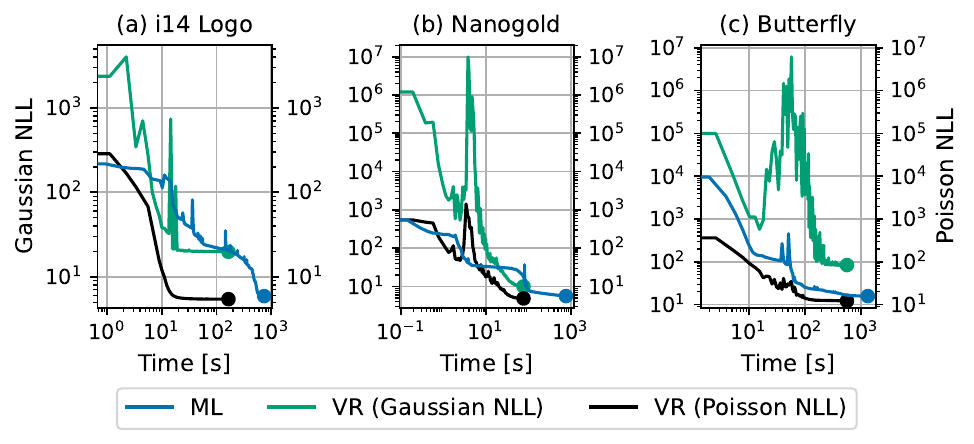}
    \caption{Loss as a function of cumulative runtime for ML and VR on the (a) nanoprobe test target, (b) Au particles and (c) butterfly scale (large) datasets. The left axis shows the Gaussian NLL, the common reporting metric; for VR it is recorded in a separate, untimed run. The right axis shows the Poisson NLL that VR minimises. Circles mark the point at which the convergence criterion is satisfied. Both axes are logarithmic.}
    \label{Experimental loss curve comparison}
\end{figure}

\paragraph{Loss Curves}
Figure~\ref{Experimental loss curve comparison} shows the Gaussian NLL as a function of cumulative runtime for ML and VR on the (a) nanoprobe test target, (b) Au particles and (c) butterfly scale (large) datasets, together with the Poisson NLL that VR minimises (right-hand axis). VR satisfies the convergence criterion first on all three datasets. Its terminal Gaussian NLL is higher than that of ML, by factors of $3.3$, $1.8$ and $5.3$ (Tab.~\ref{tab:experimental-dataset-reconstruction-statistics-ml-vr}), consistent with the different objectives of the two algorithms (Section~\ref{sec:Experimental_Setup}). On the Au particles and butterfly scale (large) datasets the Gaussian NLL of VR shows large transient spikes early in the reconstruction, reaching $10^{6}$--$10^{7}$, which are much smaller in the Poisson NLL and do not persist.

\begin{figure}[H]
    \centering
    \includegraphics[clip, width=0.8\textwidth]{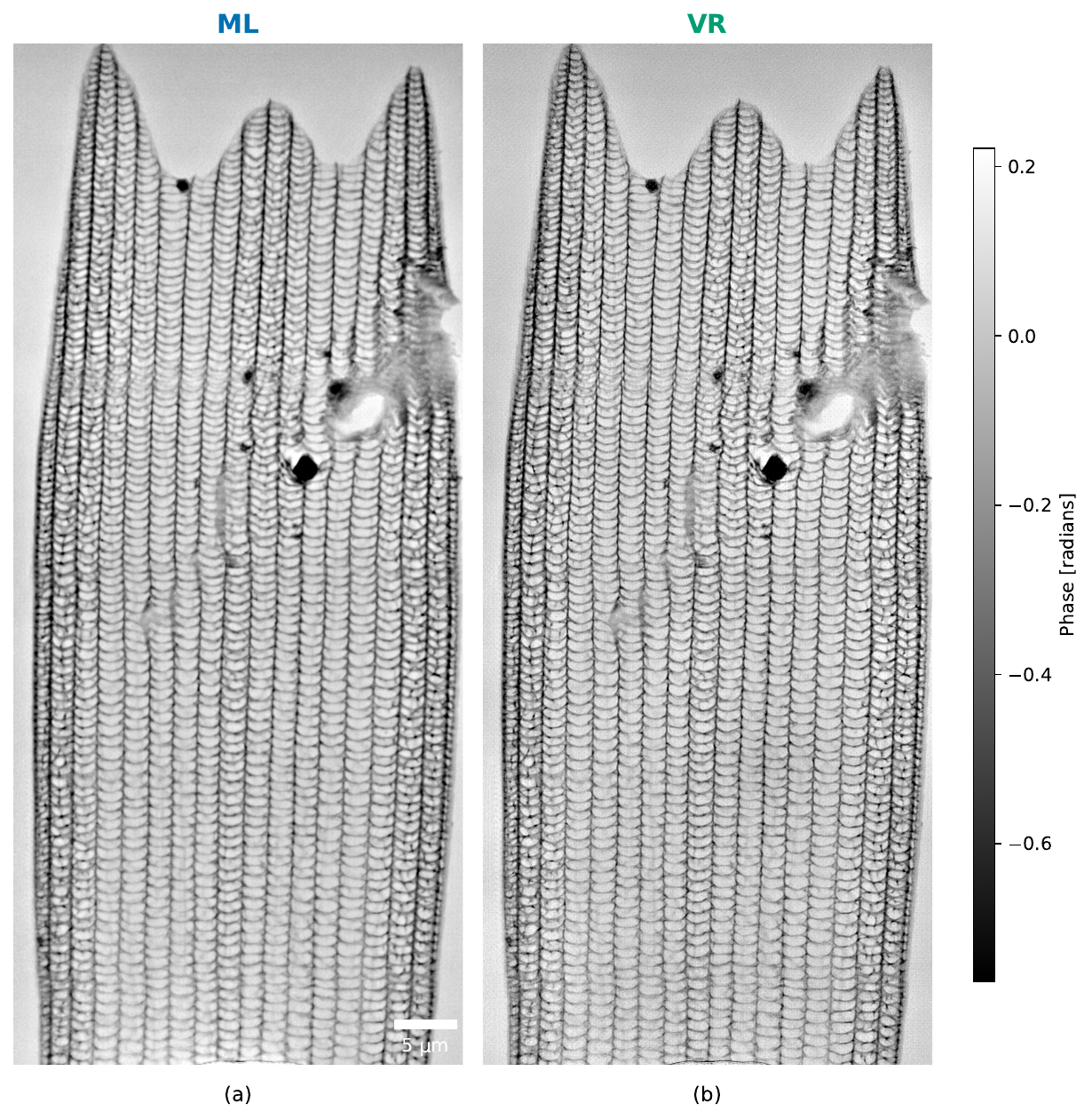}
    \caption{
Phase reconstructions of the butterfly scale large specimen obtained using (a) ML and (b) VR at the point at which the loss-plateau criterion is satisfied.
All panels share a common phase colour scale (rad). 
A spatial scale bar is shown in panel (a).
}
    \label{Butterfly object reconstruction comparison}
\end{figure}

\paragraph{Object Reconstructions}
Figure~\ref{Butterfly object reconstruction comparison} shows the butterfly scale large object reconstructions for the VR and ML algorithms at the convergence point.

\begin{figure}[H]
    \centering
    \includegraphics[clip, width=\textwidth]{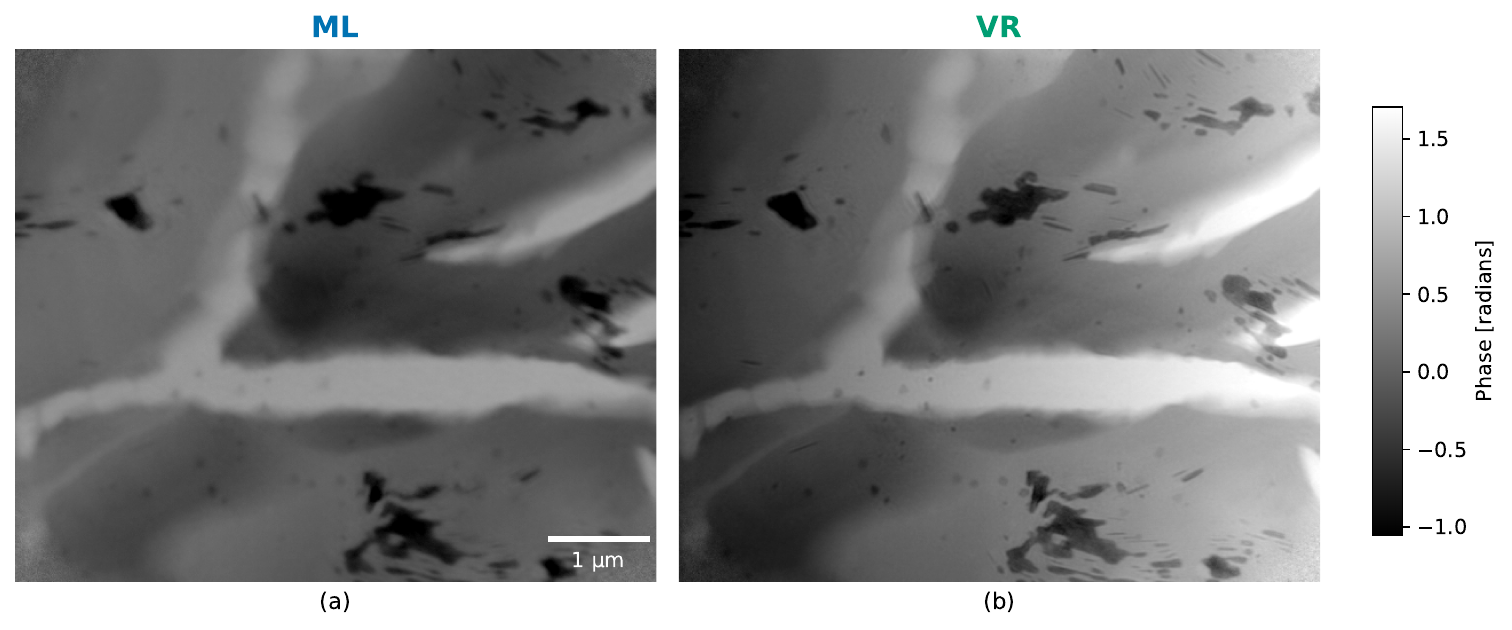}
    \caption{
Phase reconstructions of the Au particles specimen obtained using (a) ML and (b) VR at the point at which the loss-plateau criterion is satisfied.
All panels share a common phase colour scale (rad). 
A spatial scale bar is shown in panel (a).
}
    \label{Nanogold object reconstruction comparison}
\end{figure}

Figure~\ref{Nanogold object reconstruction comparison} presents the Au particles object reconstructions obtained with ML and VR at the point at which the loss-plateau criterion is satisfied. 

\begin{figure}[H]
    \centering
    \includegraphics[clip, width=\textwidth]{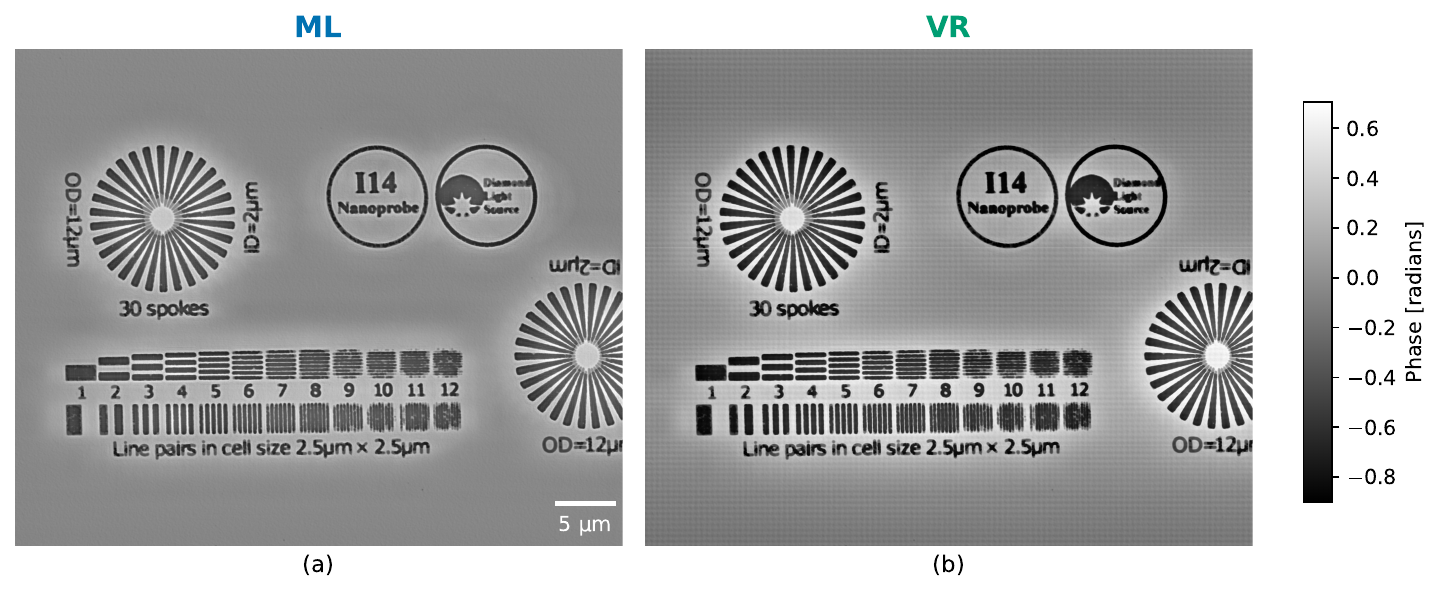}
    \caption{
Phase reconstructions of the nanoprobe test target specimen obtained using (a) ML and (b) VR at the point at which the loss-plateau criterion is satisfied.
All panels share a common phase colour scale (rad). 
A spatial scale bar is shown in panel (a).
}
    \label{i14_logo_object_reconstruction_comparison}
\end{figure}

Figure~\ref{i14_logo_object_reconstruction_comparison} shows the nanoprobe test target object reconstructions at convergence for ML and VR at the point of loss gradient covergence.

On the butterfly scale (large) dataset the ML and VR reconstructions are visually indistinguishable at the scale shown (Fig.~\ref{Butterfly object reconstruction comparison}). On the Au particles dataset VR resolves more small, high-contrast features than ML and shows more high-frequency texture (Fig.~\ref{Nanogold object reconstruction comparison}), the same trade-off between fine-scale detail and noise seen on the butterfly scale (small) dataset. On the nanoprobe test target both algorithms resolve the line-pair groups up to approximately group~8 (Fig.~\ref{i14_logo_object_reconstruction_comparison}).

\begin{figure}[H]
    \centering
    \includegraphics[width=0.8\textwidth]{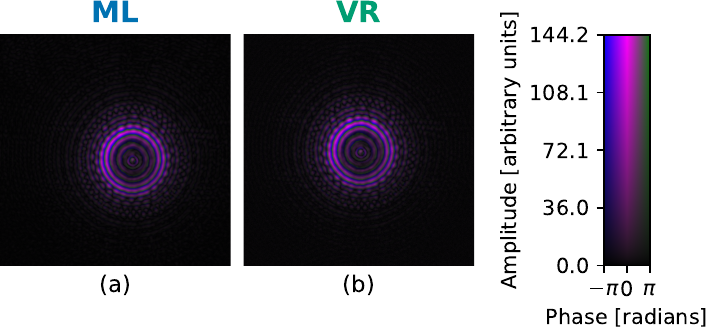}
    \caption{
Reconstructed probe amplitude and phase for the butterfly scale large dataset obtained using (a) ML and (b) VR at the point at which the loss-plateau criterion is satisfied.
Amplitude is encoded as brightness and phase as hue, according to the colour key.
}
    \label{Butterfly_probe_reconstruction_comparison}
\end{figure}

\paragraph{Probe Reconstructions}
Figure~\ref{Butterfly_probe_reconstruction_comparison} shows the reconstructed probe amplitude and phase for the butterfly scale large dataset at convergence for ML and VR. As the probe was modelled as multimodal for this dataset, the images shown in Fig.~\ref{Butterfly_probe_reconstruction_comparison} correspond to the leading singular mode obtained from the singular value decomposition of the reconstructed probe for each algorithm.

\begin{figure}[H]
    \centering
    \includegraphics[width=0.8\textwidth]{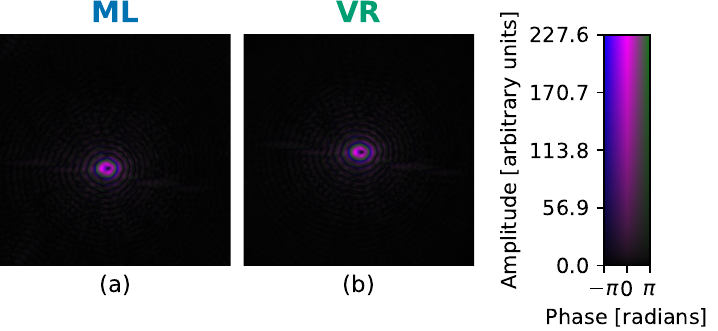}
    \caption{
Reconstructed probe amplitude and phase for the Au particles dataset obtained using (a) ML and (b) VR at the point at which the loss-plateau criterion is satisfied.
Amplitude is encoded as brightness and phase as hue, according to the colour key.
}
    \label{Nanogold_probe_reconstruction_comparison}
\end{figure}

Fig.~\ref{Nanogold_probe_reconstruction_comparison} shows the reconstructed probe amplitude and phase for the Au particles dataset. As the probe was modelled as multimodal for this dataset, the images shown in Fig.~\ref{Nanogold_probe_reconstruction_comparison} correspond to the leading singular mode obtained from the singular value decomposition of the reconstructed probe for each algorithm.

\begin{figure}[H]
    \centering
    \includegraphics[width=0.8\textwidth]{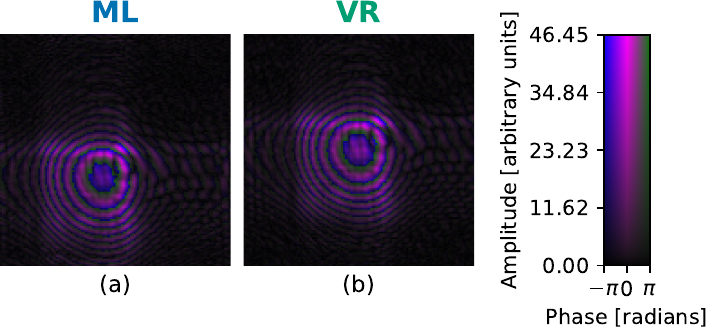}
    \caption{
Reconstructed probe amplitude and phase for the nanoprobe test target dataset obtained using (a) ML and (b) VR at the point at which the loss-plateau criterion is satisfied.
Amplitude is encoded as brightness and phase as hue, according to the colour key.
}
    \label{i14_logo_probe_reconstruction_comparison}
\end{figure}

Figure~\ref{i14_logo_probe_reconstruction_comparison} shows the reconstructed probe amplitude and phase for the nanoprobe test target dataset.

The ML and VR probes show the same structure on all three datasets but are translated relative to one another. Given that a common translation of object and probe leaves the diffraction data unchanged~\cite{Thibault_2009}, a shift in the probe reconstruction does not indicate a difference in reconstruction quality.

\paragraph{Visual Reconstruction Quality of the VR Algorithm}
Across the experimental benchmarks the variance-reduction (VR) algorithm produces object-phase reconstructions that are visually consistent with those from established methods (DM, ePIE, ML). On the butterfly scale small example, VR, DM, and ML recover comparable structural content at the stopping point of the loss-plateau criterion, whereas vanilla ePIE exhibits spatially correlated phase artefacts that are not observed in VR, DM, or ML. Overall, reconstruction fidelity is at least at parity with ML, with fewer visible artefacts than vanilla ePIE and without the mild blurring observed in the DM and ML object reconstructions.

\paragraph{Convergence Behaviour of the VR Algorithm}
Empirically, VR satisfies the loss-plateau criterion in less wall-clock time than ML. VR uses a single-stage optimisation strategy that minimises the Poisson NLL throughout; unlike ML, which minimises the Gaussian NLL.

\section{Numerical Evaluation of Computational Complexity}
\label{sec:Simulation}

This section validates the computational complexity analysis of Section~\ref{subsec:Computational_Complexity_Analysis} through numerical experiments. We first describe how the synthetic ptychographic datasets are generated: the probe positions (Section~\ref{Synthetic Probe Positions}), the forward model, and the object construction (Section~\ref{sec:synthetic_dataset_construction}). We then describe the benchmarks which measure the constants of the time expressions, and compare the peak GPU memory usage and time per pass of the VR and ML algorithms against the expressions of Section~\ref{subsec:Computational_Complexity_Analysis}.

\subsection{Generating Probe Positions for Synthetic Datasets}
\label{Synthetic Probe Positions}

Probe positions on a regular grid induce ill-posedness and lead to reconstruction artefacts~\cite{Huang_2014}. Probe positions for the synthetic datasets are generated on a nominally uniform grid with added Gaussian distributed random displacements to avoid raster-grid artefacts~\cite{Huang_2014}. For a desired overlap ratio and probe diameter, the nominal step size between adjacent positions is computed using the expressions proposed by Moshtaghpour and Kirkland~\cite{Moshtaghpour_2025}.

\subsection{Synthetic Dataset Construction}
\label{sec:synthetic_dataset_construction}

The forward model used to generate synthetic data is described by Eq.~\ref{Generator Forward Model}
\begin{equation}
\label{Generator Forward Model}
\boldsymbol{I}_{k}
=
Po\!\left( \kappa_k
\left| \mathcal{F}\!\left[
\boldsymbol{P}(\boldsymbol{r}-\boldsymbol{r}_k) \cdot \boldsymbol{O}
\right]
\right|^2
\right),
\end{equation}
where $Po(\cdot)$ is applied element-wise, with each output sampled from a Poisson distribution whose mean is given by the argument. The propagation operator is approximated by the discrete Fourier transform, $\mathcal{F}$. The parameter $\kappa_k \in \mathbb{R}^1_+$ is a scalar intensity-scaling factor used to simulate probe variation between diffraction patterns, as commonly observed in experimental acquisitions.

\begin{figure}[H]
    \centering
    \includegraphics[clip, width=0.8\textwidth]{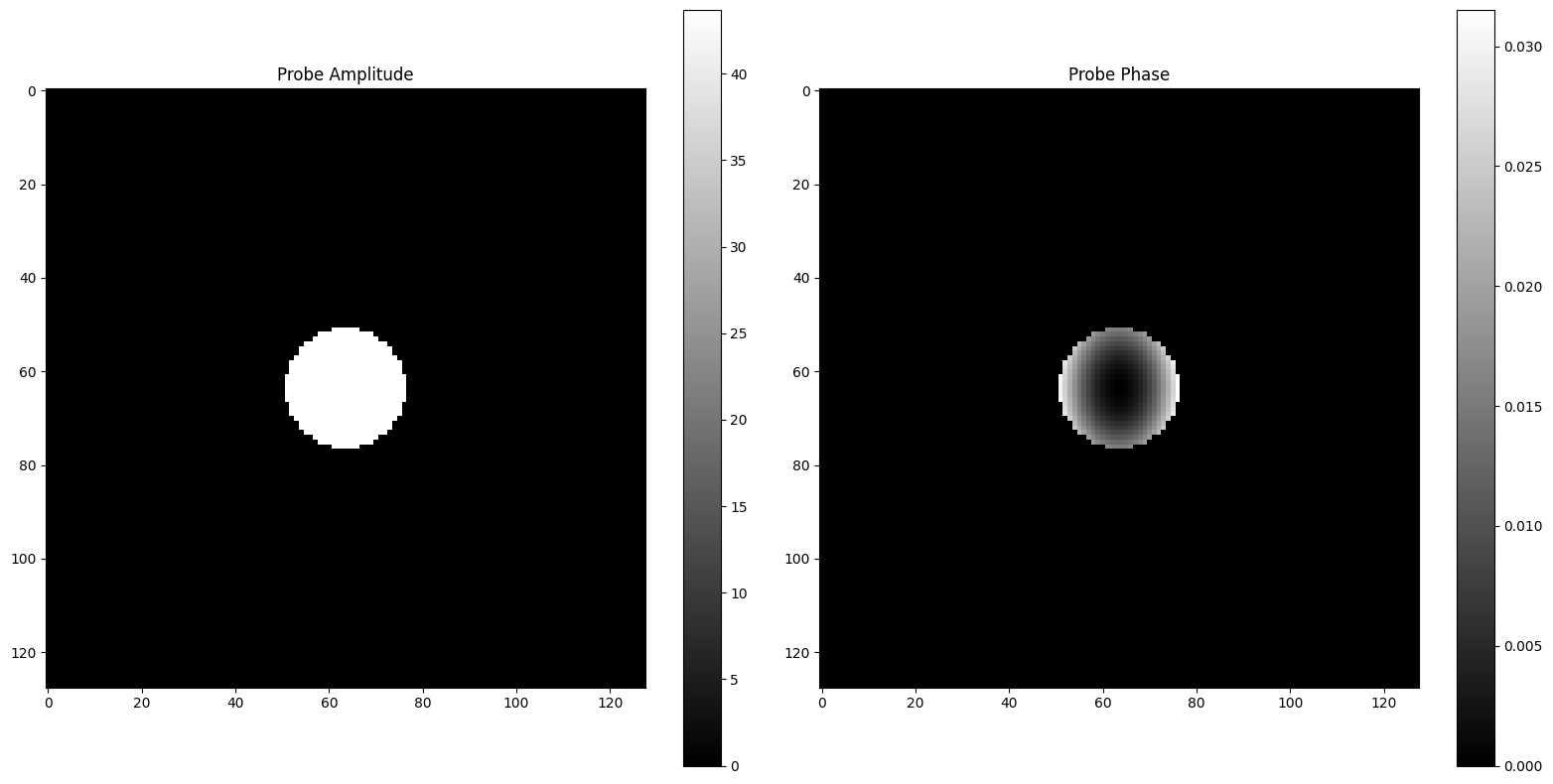}
    \caption{Amplitude (left) and phase (right) of the complex-valued probe used to generate the synthetic ptychographic datasets.}
    \label{Synthetic Ptychography Probe}
\end{figure}

The probe is simulated as a disc in both phase and amplitude, with added coma, astigmatism, and defocus. The model probe used during reconstruction is identical to the probe used to generate the data, up to a small random perturbation in the added aberration terms. The probe used to generate the synthetic dataset is shown in Figure~\ref{Synthetic Ptychography Probe}.

\begin{figure}[H]
    \centering
    \includegraphics[clip, width=\textwidth]{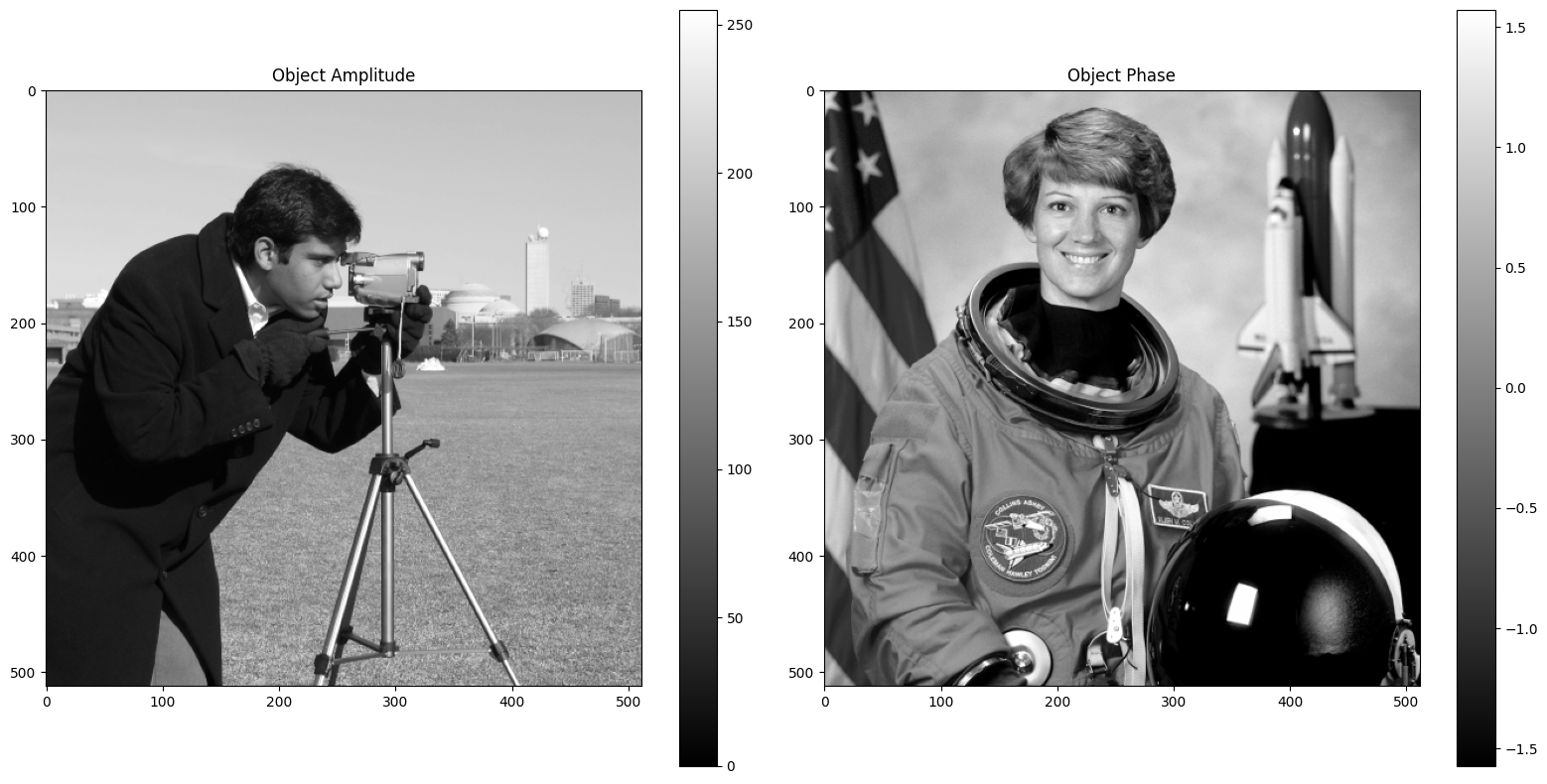}
    \caption{Amplitude (left) and phase (right) of the complex-valued synthetic object used to generate the diffraction data for the scaling experiments.}
    \label{Synthetic Ptychography Object Tile}
\end{figure}

Similar to other simulated ptychography demonstrations~\cite{Maiden_2009, Nashed_2014}, the synthetic object is constructed from two greyscale stock images of size $512 \times 512$ pixels drawn from the \texttt{skimage} Python library~\cite{vanderWalt_2014}. The \texttt{astronaut} image defines the object phase, while the \texttt{camera} image defines the object magnitude.  The complex-valued object formed from the \texttt{astronaut} phase image and the \texttt{camera} magnitude image is shown in Fig.~\ref{Synthetic Ptychography Object Tile}.

To generate larger datasets with the same overlap ratio, the $512\times512$ base object is tiled in each dimension. This ensures that larger datasets consist of repeated structure. Holding the underlying object fixed while varying only the number of scan positions and diffraction patterns isolates the effect of dataset size on scaling behaviour.

To empirically validate the memory and time expressions of Section~\ref{subsec:Computational_Complexity_Analysis}, synthetic datasets are constructed in which only the number of diffraction patterns is varied. The dimensions of these synthetic datasets are given in Table~\ref{Synthetic Datasets Table}. All diffraction patterns have dimensions $128 \times 128$ pixels.

\begin{table}[H]
\centering
\caption{Synthetic datasets used for computational scaling validation. The number of diffraction patterns and reconstructed object grid size are increased systematically to assess algorithmic scaling behaviour. The mini-batch size $K'$ of each VR reconstruction is stated, the corresponding \texttt{PtyPy} block size is $F = 6{,}000$ patterns for every dataset.}
\label{Synthetic Datasets Table}
\begin{tabular}{|c|c|c|}
\hline
\textbf{Number of diffraction patterns} & \textbf{Object size} & \textbf{Mini-batch size $K'$} \\
\hline
$1.52 \times 10^3$ & $511 \times 510$ & $64$ \\
\hline
$1.99 \times 10^4$ & $1,530 \times 1,530$ & $1,024$ \\
\hline
$1.20 \times 10^5$ & $3,581 \times 3,581$ & $4,096$ \\
\hline
$1.02 \times 10^6$ & $10,240 \times 10,240$ & $16,384$ \\
\hline
\end{tabular}
\end{table}

\subsection{Scaling Results}
\label{sec:Scaling results}

\paragraph{Measurement of the Time Constants}
The constants of Eqs.~\ref{VR time equation} and~\ref{ML time equation} are obtained from benchmarks that time single reconstruction steps of the code, optimiser settings and \texttt{PtyPy} parameter tree used in the reconstructions below. For VR, a sweep of the mini-batch size from $64$ to $16{,}384$ patterns at a fixed $3{,}581 \times 3{,}581$ object gives $a_{\mathrm{VR}}$ as the slope of the step time against $K' N_{\mathrm{probe}}$; a sweep of the object dimension from $511$ to $10{,}240$ pixels at a fixed mini-batch of $1{,}024$ patterns gives $b_{\mathrm{VR}}$ as the slope against $N_{\mathrm{params}}$; and $g_{\mathrm{VR}}$ is the intercept of either sweep. For ML, individual iterations are timed on seven synthetic datasets of between $1.5 \times 10^{3}$ and $2.5 \times 10^{5}$ patterns, generated as in Section~\ref{sec:synthetic_dataset_construction}, with $a_{\mathrm{ML}}$ and $g_{\mathrm{ML}}$ the slope and intercept against $K N_{\mathrm{probe}}$. Each step is timed after warm-up with device synchronisation, and the median of $20$ steps (VR) or $50$ iterations (ML) is used.  

\begin{figure}[H]
    \centering
    \includegraphics[clip, width=\textwidth]{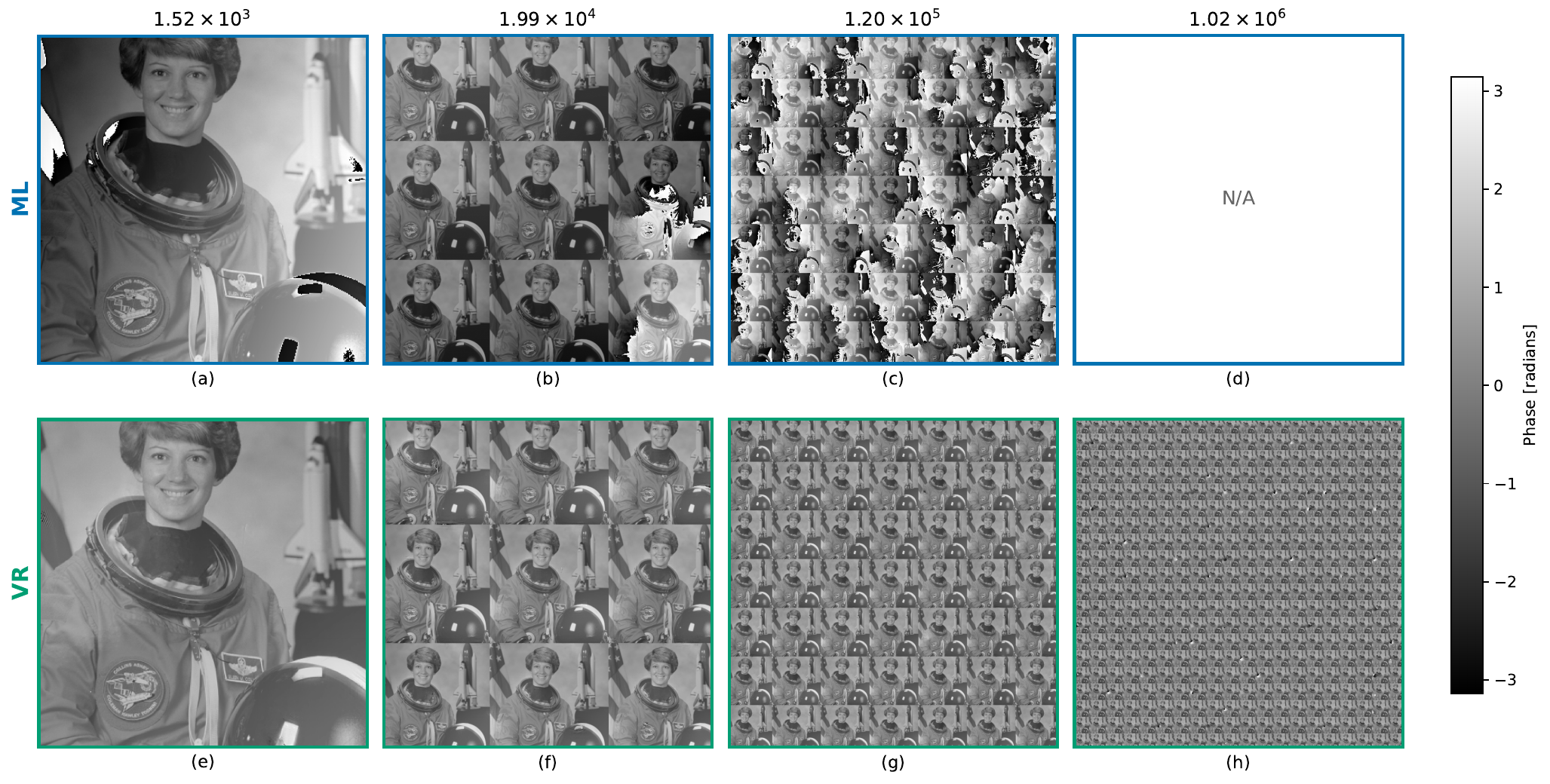}
    \caption{
Phase reconstructions of the synthetic object obtained using ML, panels (a)--(d), and VR, panels (e)--(h), at the point at which the loss-plateau criterion is satisfied. Each column corresponds to one of the synthetic datasets of Tab.~\ref{Synthetic Datasets Table}, containing (a, e) $1.52 \times 10^{3}$, (b, f) $1.99 \times 10^{4}$, (c, g) $1.20 \times 10^{5}$, and (d, h) $1.02 \times 10^{6}$ diffraction patterns. Panel (d) is marked N/A because the ML algorithm was not run on the largest dataset, for the reason given in Section~\ref{sec:Scaling results}. Neither the constant phase offset nor the linear phase ramp of a reconstruction is determined by the data; both are removed from a panel where doing so reduces the wrapping of the displayed phase, and reconstructions that do not wrap are displayed as reconstructed. All panels share a common phase colour scale given in radians.}
    \label{Synthetic object reconstruction comparison}
\end{figure}

\paragraph{Synthetic Object Reconstructions} Figure~\ref{Synthetic object reconstruction comparison} shows the object phase reconstructions obtained with ML and VR for each of the synthetic datasets of Tab.~\ref{Synthetic Datasets Table}. Both algorithms recover the tiled object at every dataset size at which they can be executed, so the scaling measurements that follow are obtained from reconstructions of comparable structural content. For the two largest datasets on which ML is feasible, the ML reconstructions show phase offsets between tiles of the object that are large enough for the recovered phase to wrap; the VR reconstructions remain within a single phase branch at all four dataset sizes.

Table~\ref{tab:synthetic-time-summary} shows the time per iteration over the full dataset, the number of iterations to convergence and the resulting time to convergence for both ML and VR across increasing numbers of diffraction patterns, with the time per iteration compared against the values predicted by Eqs.~\ref{VR time equation} and~\ref{ML time equation}.

\begin{table}[H]
\centering
\caption{Time per pass over the data for the synthetic datasets: one full-batch iteration for ML and one epoch for VR. Measured values are the reported time to convergence divided by the number of passes $L$; predicted values are Eqs.~\ref{VR time equation} and~\ref{ML time equation} with the constants of Tab.~\ref{tab:time-constants}.}
\label{tab:synthetic-time-summary}
\footnotesize
\begin{tabular}{|c|c|c|c|c|c|c|c|c|}
\hline
\multirow{2}{*}{\shortstack{\textbf{Number of}\\\textbf{patterns}}} & \multicolumn{4}{c|}{\textbf{ML}} & \multicolumn{4}{c|}{\textbf{VR}} \\ \cline{2-9}
 & \shortstack{$t_{\mathrm{iter}}$ (s)\\measured} & \shortstack{$t_{\mathrm{iter}}$ (s)\\predicted} & $L$ & \shortstack{$T$ (s)\\measured} & \shortstack{$t_{\mathrm{epoch}}$ (s)\\measured} & \shortstack{$t_{\mathrm{epoch}}$ (s)\\predicted} & $L$ & \shortstack{$T$ (s)\\measured} \\ \hline
$1.52 \times 10^3$ & 0.0080 & 0.0105 ($+30.6\%$) & 5855 & 47.0 & 0.0241 & 0.0173 ($-28.3\%$) & 266 & 6.4 \\ \hline
$1.99 \times 10^4$ & 0.104 & 0.0963 ($-7.8\%$) & 11630 & 1214.7 & 0.0923 & 0.0926 ($+0.3\%$) & 315 & 29.1 \\ \hline
$1.20 \times 10^5$ & 0.575 & 0.563 ($-2.0\%$) & 11169 & 6418.4 & 0.538 & 0.523 ($-2.8\%$) & 338 & 181.9 \\ \hline
$1.02 \times 10^6$ & N/A & N/A & N/A & N/A & 4.706 & 4.587 ($-2.5\%$) & 405 & 1906.0 \\ \hline
\end{tabular}
\end{table}

Table~\ref{tab:synthetic-memory-summary} shows the peak GPU memory, paired with the byte counts of Eq.~\ref{VR memory equation} and Eq.~\ref{ML memory equation}. Convergence is defined using the gradient-based criterion of Section~\ref{Convergence Criteria}. Peak GPU memory is the memory usage predicted during iterating each reconstruction algorithm, not accounting for any allocations of ML to GPU during data loading. For all dataset sizes where both methods were run, VR reduces peak GPU memory usage and time to convergence relative to ML.

\begin{table}[H]
\centering
\caption{Peak GPU memory of the synthetic reconstructions, measured per process by NVML over a window that spans data loading and the reconstruction, against the byte counts of Eqs.~\ref{VR memory equation} and~\ref{ML memory equation}.}
\label{tab:synthetic-memory-summary}
\begin{tabular}{|c|c|c|c|c|}
\hline
\multirow{2}{*}{\shortstack{\textbf{Number of}\\\textbf{diffraction patterns}}} & \multicolumn{2}{c|}{\textbf{ML peak memory (GB)}} & \multicolumn{2}{c|}{\textbf{VR peak memory (GB)}} \\ \cline{2-5}
 & measured & predicted & measured & predicted \\ \hline
$1.52 \times 10^3$ & 1.68 & 1.66 ($-1.5\%$) & 0.94 & 0.82 ($-12.6\%$) \\ \hline
$1.99 \times 10^4$ & 7.33 & 7.31 ($-0.3\%$) & 3.27 & 3.04 ($-7.1\%$) \\ \hline
$1.20 \times 10^5$ & 20.30 & 20.23 ($-0.4\%$) & 13.64 & 12.95 ($-5.1\%$) \\ \hline
$1.02 \times 10^6$ & N/A & 141.92 & 88.94 & 87.33 ($-1.8\%$) \\ \hline
\end{tabular}
\end{table}

‘N/A’ indicates that the ML algorithm was not run on the dataset containing $1.02 \times 10^6$ diffraction patterns. Eq.~\ref{ML memory equation} gives a resident requirement for ML of \SynDMLDataPred~GB for the diffraction data alone at the block size used here, and a predicted total of \SynDMLMemPred~GB, which exceeds the memory of a single Nvidia H100.

The benchmark constants predict the time per pass over the data in Tab.~\ref{tab:synthetic-time-summary} to within $8\%$ for every run containing $1.99\times10^{4}$ or more patterns, a factor-of-$50$ range in $K$, and to within $3\%$ for both algorithms on the two largest datasets. The largest error in this range is the ML prediction at $1.99\times10^{4}$ patterns, which is $7.8\%$ below the measured value. On the smallest dataset the errors are larger, $+31\%$ for ML and $-28\%$ for VR. This variability therefore affects the VR time to convergence, but not the time per epoch.

The byte count predicted by Eq.~\ref{ML memory equation} agrees with the measured ML peak to within $1.5\%$ at all three dataset sizes. Equation~\ref{VR memory equation} under-predicts the measured VR peak by $2\%$--$13\%$, with the discrepancy decreasing as the dataset size increases. Overall, the measured improvement factors are $7$--$42\times$ in time to convergence and $1.5$--$2.2\times$ in peak GPU memory.

The memory advantage of VR arises from the three effects identified in Section~\ref{subsec:Computational_Complexity_Analysis}: VR stores only the measured intensities, whereas \texttt{PtyPy} stores both intensities and per-pattern weights; \texttt{PtyPy} rounds its resident data allocation up to a whole number of blocks; and the transient working set of VR scales with $K'$, rather than with the block size $F$.

\section{Conclusion}
\label{sec:Conclusion}

Our method applies mini-batch stochastic optimisation with a decaying step size to the Poisson maximum-likelihood formulation of ptychography~\cite{Thibault_2012}. On the datasets tested, it produces reconstructions visually comparable to those of DM, ePIE and ML, and it satisfies the convergence criterion faster than ML with lower peak GPU memory at the configurations compared.

The transient memory of VR is set by its mini-batch size $K'$, and that of \texttt{PtyPy} ML by its block size $F$; both are user-selected parameters, so measured memory ratios depend on the configurations compared. At the settings used in this work, peak GPU memory for VR is lower than for ML on every experimental dataset, by factors of between $1.5$ and $10.8$. The largest reductions, from \MLMemory~GB to \VRMemory~GB on the butterfly scale (small) dataset and a factor of $10.8$ on the Au particles dataset, arise where ML was run with blocks far larger than the VR mini-batch ($F = 2500$ against $K' = 64$, and $F = 900$ against $K' = 16$), so that the working set of a single block dominates the ML footprint. Eq.~\ref{ML memory equation} predicts that reducing $F$ to the VR mini-batch size would lower the ML footprint on these datasets to approximately $6.6$~GB and $2.9$~GB, within a factor of $1.5$ of the corresponding VR predictions; the effect of such small blocks on the ML runtime was not measured. The part of the advantage that is independent of $F$ and $K'$ lies in the resident data: VR stores only the measured intensities, whereas \texttt{PtyPy} additionally stores per-pattern weights, which bounds the memory advantage of VR at a factor of two once the resident data dominates, as for the nanoprobe test target ($1.5\times$). \texttt{PtyPy} can also reduce its device footprint by streaming blocks between host and device, at a transfer cost not captured by Eq.~\ref{ML memory equation}; the largest synthetic dataset is therefore infeasible for ML on a single H100 only without such streaming.

Synthetic scaling experiments, in which only the number of diffraction patterns is varied, confirm the complexity analysis: with constants measured independently of the reconstructions, the time per pass is predicted to within $8\%$ for every run of more than $10^{4}$ patterns, and to within $3\%$ on the two largest datasets, the ML peak memory to within $1.5\%$, and the VR peak memory to within $13\%$, over three orders of magnitude in frame count. Peak memory grows linearly with frame count for both algorithms, but with half the slope for VR, and the time to convergence of VR grows more slowly than that of ML.

Taken together, these results show that VR achieves a lower memory footprint than \texttt{PtyPy} ML at the configurations tested, with an advantage of up to a factor of two in resident data that persists for any block size, and favourable computational scaling for large-scale ptychographic reconstruction.

From a broader computational imaging perspective, this work illustrates the practical value of approximation techniques from the machine learning literature—such as stochastic gradient descent and variance reduction—when applied to large-scale inverse problems. While SGD and its variants only approximate the full gradient, on the datasets tested the associated approximation error does not visibly degrade the reconstruction relative to full-batch ML. Whether mini-batch gradient noise also acts as an implicit regulariser in ptychography, as has been argued for deep learning~\cite{Hoffer_2017}, is not tested here. The stochastic gradient noise introduced by mini-batch sampling can, in turn, be tuned via the mini-batch size.

The present work focuses on a Poisson maximum-likelihood formulation of ptychography. Extending variance-reduced optimisation to alternative formulations, including set-projection algorithms, remains an open direction. Since only the mini-batch size and the step-size decay factor were set for each dataset, and since similar algorithms are used across different ptychographic modalities, it is reasonable to expect that the proposed variance-reduced reconstruction framework would transfer with limited modification to other modalities such as electron ptychography.

The argument that variance-reduced optimisation should transfer effectively to related imaging modalities also applies to higher-dimensional imaging, such as ptycho-tomography~\cite{Grsoy_2017,Dierolf_2010}, where the dimensionality of the inverse problem is substantially larger. In higher-dimensional imaging settings such as ptycho-tomography, the scaling advantages of stochastic and variance-reduced methods over full-gradient optimisation may be even more pronounced.

Further computational gains may be attainable by employing adaptive optimisers with lower memory requirements than Adam. Examples include MicroAdam~\cite{Modoranu_2024}, a memory-efficient optimiser that compresses gradient information, and Adafactor~\cite{Shazeer_2018}, which reduces memory requirements by factorising second-moment estimates. Optimisers such as Shampoo~\cite{Gupta_2018}, which have demonstrated faster convergence than Adam on certain machine learning tasks, could also be considered. In this work, we restrict ourselves to established optimisers available within the \texttt{PyTorch} ecosystem; evaluating alternative optimisers in the context of ptychographic reconstruction, and quantifying their impact on convergence, memory usage, and hyperparameter sensitivity, is left for future work.

An important avenue for improvement is the development of more informative reconstruction quality metrics, which would better indicate both convergence and solution quality~\cite{Thibault_2024}. Designing such metrics is challenging, as ptychography is an ill-posed inverse problem without a unique ground-truth solution.

Finally, there is an opportunity to further explore connections between generalisation in deep learning and robustness in inverse problems. In supervised learning, hyper-parameters such as batch size influence the generalization gap between training and unseen data~\cite{Hoffer_2017,Li_2014}. In ptychography and related inverse problems, analogous hyper-parameters affect the ability of an optimisation algorithm to fit an imperfect forward model to experimental data while avoiding over-fitting to noise or artefacts. Concepts such as small-batch training and implicit regularisation may therefore translate between these domains, which motivates deeper theoretical and empirical study of the relationship between generalization in deep learning and robustness in inverse problems.

\begin{acknowledgements}
This work was carried out with the support of Diamond Light Source, instruments I14 (in-house data) and I08-1 (proposal MG36964 and in-house data). The authors would like to thank Dr. Duane Loh, Dr. Anupama Prakash, Mr. Ervin Chia and Mr. Justin Ong for their help and support during the MG36964-1 experiment. The authors would further like to thank Dr. Majid Kazemian for the in-house data collection of the Au nanoparticles dataset. The authors would also like to thank Dr. Miguel Gomez Gonzalez for the in-house data collection of the I14 nanoprobe test target.  
\end{acknowledgements}

\begin{funding}
A.W.M. was supported by a Scientific Computing PhD studentship from the Ada Lovelace Centre.

J.T. is supported by the UK Engineering and Physical Sciences Research Council (EPSRC) through the grant EP/Y028872/1; Mathematical Foundations of Intelligence: An “Erlangen Programme” for AI.
\end{funding}

\ConflictsOfInterest{There are no conflicts of interest.
}

\DataAvailability{
The Au nanoparticles and nanoprobe test target datasets are both available here: https://doi.org/10.5281/zenodo.11501765. The butterfly scale dataset is available upon reasonable request. 

The implementation of the variance-reduced reconstruction framework, together with the scripts used to perform all reconstructions and generate all figures and tables reported in this article, will be made openly available at https://github.com/am318/variance-reduction-ptychography after publication.
}

\bibliography{References}

\end{document}